# SCOTCH – Search for Clandestine Optically Thick Compact HII regions: II

A. L. Patel,[1★] J. S. Urquhart,[1†] A. Y. Yang,[2,3‡] T. Moore,[4] M. A. Thompson,[5] K. M. Menten,[6] T. Csengeri,[7]

[1] *Centre for Astrophysics and Planetary Science, University of Kent, Canterbury, CT2 7NH, UK*
[2] *National Astronomical Observatories, Chinese Academy of Sciences, A20 Datun Road, Chaoyang District, Beijing 100101, P. R. China*
[3] *Key Laboratory of Radio Astronomy and Technology, Chinese Academy of Sciences, A20 Datun Road, Chaoyang District, Beijing, 100101, P.R. China*
[4] *Astrophysics Research Institute, Liverpool John Moores University, Liverpool Science Park, 146 Brownlow Hill, Liverpool, L3 5RF, UK*
[5] *School of Physics and Astronomy, University of Leeds, Leeds LS2 9JT, UK*
[6] *Max Planck Institute for Radio Astronomy, Auf dem Hügel 69, 53121 Bonn, Germany*
[7] *Laboratoire d'astrophysique de Bordeaux, Univ. Bordeaux, CNRS, B18N, allée Geoffroy Saint-Hilaire, F-33615 Pessac, France*

Accepted XXX. Received YYY; in original form ZZZ

**ABSTRACT**

In this study we present 18−24 GHz and high-angular-resolution (0.5 arcsec) radio wavelength Australia Telescope Compact Array (ATCA) follow-up observations towards a sample of 39 HC HII region candidates. These objects, taken from a sample hosting 6.7 GHz methanol masers, were chosen due to the compact and optically thick nature of their continuum emission. We have detected 27 compact radio sources and constructed their spectral energy distributions (SEDs) over the 5–24 GHz range to determine the young HII region's physical properties, i.e., diameter, electron density $n_e$, emission measure, Lyman continuum flux $N_{Ly}$ and turnover frequency $\nu_t$. The flux measurements are fitted for 20 objects assuming an ionisation-bounded HII region with uniform density model. For the remaining 7 objects that lack constraints spanning both their optically thick and thin regimes, we utilise relations from the literature to determine their physical properties. Comparing these determined parameters with those of known hypercompact (HC) and ultracompact (UC) HII regions, we have identified 13 HC HII regions, 6 intermediate objects that fall between HC HII and UC HII regions, 6 UC HII regions and one radio jet candidate which increases the known population of HC HII regions by ∼50 per cent. All the young and compact HII regions are embedded in dusty and dense clumps and ∼80 per cent of the HC HII regions identified in this work are associated with various maser species ($CH_3OH$, $H_2O$ and OH). Four of our radio sources remain optically thick at 24 GHz; we consider these to be amongst the youngest HC HII regions.

## 1 INTRODUCTION

Massive stars (> 8 M$_\odot$, > 10$^4$ L$_\odot$) generate intense Lyman continuum emissions with sufficient energy to ionize their surroundings, resulting in the formation of HII regions. These hot bubbles of dense ionized gas are over-pressured with respect to their surrounding medium and hence expand over time. Their expansion drives ionisation shocks that propagate into the ambient medium (Dyson et al. 1995). The smallest and most compact HII regions are naturally assumed to be the youngest, however, many details of their evolution and formation remain unclear. An important question that remains unresolved in the formation of HII regions, is how accretion proceeds against the outward pressure within the ionisation front.

Many theoretical models have predicted the outcome for the formation and evolution of HII regions. For example, the McKee & Tan (2003) turbulent core and Peters et al. (2010) ionisation feedback models suggest that the youngest HII regions expand into outflow driven cavities away from the disk accretion flows. This implies that the expansion of young HII regions does not halt the accretion flow onto young high-mass star. A similar result has been reported in Keto (2007) who suggests that while the young high-mass star grows, emitting an increasing amount of ionizing UV photons, the surrounding HII region goes through a 3-stage evolutionary sequence: non-existent/quenched, bipolar outflow and spherical outflows. Contrary to this, the models of Hosokawa & Omukai (2009); Hosokawa et al. (2010) suggest the high accretion rates of material falling onto massive stars cause the protostar to bloat and swell (and pulsate). This reduces the effective temperature of the star and delays the development of an Hii region. It is clear that the main differences in these theoretical models arise at the initial stage of Hii region development. Investigating the physical characteristics of the youngest and smallest Hii regions is crucial to understand their initial conditions and constrain theoretical models.

★ E-mail: alp48@kent.ac.uk
† E-mail: j.s.urquhart@kent.ac.uk
‡ E-mail: yangay@nao.cas.cn

The two youngest Hii region stages are commonly known as hypercompact Hii (HC Hii) and ultracompact (UC Hii) Hii regions (Kurtz & Hofner 2005). HC Hii regions are the youngest and are defined by the following physical parameters: physical diameter ≤ 0.05 pc, an electron density ($n_e$) of ≥ $10^5$ cm$^{-3}$, an emission measure of ≥ $10^8$ pc cm$^{-6}$. The brightest HC Hii regions are associated with radio recombination line (RRL) linewidths of ΔV ≥ 40 km s$^{-1}$ (Kurtz et al. 1999; Sewilo et al. 2004; Hoare et al. 2007; Murphy et al. 2010). The next evolutionary phase is the UC Hii region; these have typical diameters of ≤ 0.1 pc, an $n_e$ of ≥ $10^4$ cm$^{-3}$, an emission measure of ≥ $10^7$ pc cm$^{-6}$, and RRL linewidths ΔV ≥ ∼25-30 km s$^{-1}$ (Wood & Churchwell 1989; Afflerbach et al. 1996; Hoare et al. 2007).

HC Hii regions are particularly interesting in the study of massive star formation because they are thought to exhibit a size scale suitable for a single high-mass star (or a binary system), in contrast to ultracompact (UC) or compact Hii regions, which are more likely to harbour clusters of massive stars (e.g. Kurtz 2000; Hoare et al. 2007). Theoretical models suggests that O-type stars form gradually via accretion through an HC Hii region onto intermediate-mass B stars (Keto 2007). However, this has yet to be definitively confirmed. Thus, the study of HC Hii regions is important to understand the evolution of massive stars. The differences between these two evolutionary classes are somewhat arbitrary as the evolution from the HC Hii region stage to the UC Hii region stage is thought to be continuous (Garay & Lizano 1999; Yang et al. 2019). Currently the number of known HC Hii regions is small and a larger sample with good statistics is needed to constrain the properties of this very early stage of massive star formation.

To date, only 23 HC Hii regions have been identified based on their emission nature at centimetre, infrared (IR) and sub-millimetre wavelengths (Yang et al. 2019; Yang et al. 2021 and references therein). These HC Hii region were selected based on their rising spectral index at 5 GHz. Young Hii regions are deeply embedded and heavily obscured by thick cocoons of molecular gas that are opaque in the optical and near-infrared regimes; however, they are transparent to radio emission. We can therefore use radio-continuum observations to detect these regions and characterise their properties. The objective of this study is to investigate and confirm the nature of a sample of early-stage Hii regions identified in Patel et al. (2023); (hereafter Paper I) and is specifically aimed at increasing the sample of HC Hii regions.

In Paper I, we have presented a detailed analysis of high-frequency (i.e., 23.7-GHz) radio-continuum emission towards 141 sources associated with 6.7-GHz class II methanol masers, as identified by the Methanol MultiBeam survey (MMB; Green et al. 2009). Paper I used archival data taken with the Australia Telescope Compact Array by Titmarsh et al. (2016). Class II Methanol masers are known to be exclusively associated with the early stages of high-mass star formation (Breen et al. 2013), and are therefore an ideal place to search for HC Hii regions. This low-resolution (∼20-arcsec) study detected 68 compact radio sources. We conducted a multi-wavelength analysis of these objects and identified 49 Hii regions, of which, the emission of 13 are optically thick between 5 and 23.7 GHz.

In this paper we report the results of high angular resolution (∼0.5-arcsec) observations towards the sample of 39 compact (i.e., y-factor ($f_{\rm int}/f_{\rm peak}$) < 2)[1] and/or optically thick ($\alpha$ > 0.5) Hii regions identified in Paper I. We derive the physical properties for each radio source and classify their properties into three types: HC Hii regions, UC Hii regions, and intermediate transition objects that are between the HC Hii and UC Hii regions stages. The structure of the paper is as follows: the observations, data reduction, imaging and source extraction methods are described in Section 2. Section 3 presents the results, a discussion of the physical parameters and the methods used to derive them. In Section 4, we investigate the evolution of Hii regions by comparing our sample with known compact Hii regions. We divide our Hii regions into sub classes based on their physical properties and discuss the youngest Hii regions within our sample. We summarise our results and highlight our main findings in Section 5.

## 2 OBSERVATIONS AND DATA REDUCTION

### 2.1 Observational setup

The Australia Telescope Compact Array (ATCA) is an array of six 22-m antennas located 500 km north-west of Sydney, in the Paul Wild Observatory. Five of these antennas can move along a 3-km long east-west track or a 214-m north spur. The sixth antenna is fixed, being located out 6 km along the east-west track. The ATCA is an Earth-rotation aperture-synthesis radio interferometer and has 17 standard configurations that are designed to give optimum and minimum-redundancy coverage of the visibilities ($u, v$) plane after a 12-hour observing period. The observations presented here were collected over four days in April 2022 towards 39 fields between Galactic longitudes 2°– 18° and 326°– 340°. These fields were selected towards radio sources that are either found unresolved or optically thick at 23.7 GHz based on the low-resolution observations (20 arcsec) presented in Paper I.

The observations were made using the 6-km configuration (6A), utilising all six antennae in a linear orientation with the shortest and longest baselines being 337 and 5939 m, respectively. The Compact Array Broad-band Backend (CABB; Wilson et al. 2011) was configured with 2 × 2-GHz continuum bands with 32 × 64-MHz channels to observe at 18 and 24 GHz simultaneously. The primary beam for the ATCA has a full width half maximum (FWHM) size of 2.6 arcmin and the synthesised beam of the 6A antenna configuration is ∼ 0.4 arcsec (FWHM) at 24 GHz. We simultaneously measured 5 zoom windows at 17.3, 18.0, 18.7, 23.4, 24.5 GHz targeting the H72$\alpha$, H71$\alpha$, H70$\alpha$, H65$\alpha$ and H64$\alpha$ radio recombination lines (RRLs), respectively (Brown et al. 1978). Here we present the results obtained from the 2 GHz continuum bands centered at 18 and 24 GHz, the results from the RRL observations will be presented in a future paper. Target sources were divided into four blocks based on their position and observations of each field typically consisted of six 1.5-minute integration cuts, that were spaced over a range of hour angles to maximise the $uv$ coverage. The total on-source integration time of ∼ ten minutes gives a theoretical rms noise of ∼0.1 mJy/beam at both frequencies, which is sufficient to detect a B0.5 or earlier type star embedded within an ionised nebula at a distance of 20 kpc (Giveon et al. 2005; Urquhart et al. 2007b).

The observations were generally carried out over a 10 hr period to provide even $uv$-coverage, however, for higher declination sources (i.e., close to the celestial equator $l = 2° - 18°$) the observations were limited to a 5 hr period. The synthesised beams for sources at higher declination are relatively poor given that the ATCA is an E-W interferometer. Consequently, the major axis of the beam is significantly elongated (average aspect ratio ∼ 5:1). Table 1 provides the theoretical size of the beam and RMS of each group based on the average declination at 18 and 24 GHz, respectively. To correct for fluctuations in the phase and amplitude caused by atmospheric conditions, the sources were sandwiched between observations of a nearby phase calibrators. The primary (1934−648) and bandpass (1252−055) calibrator were observed once during each set of observation for approximately 10 mins each, to allow the absolute calibration of the flux density and bandpass.

[1] The y-factor is the ratio of integrated to peak flux density

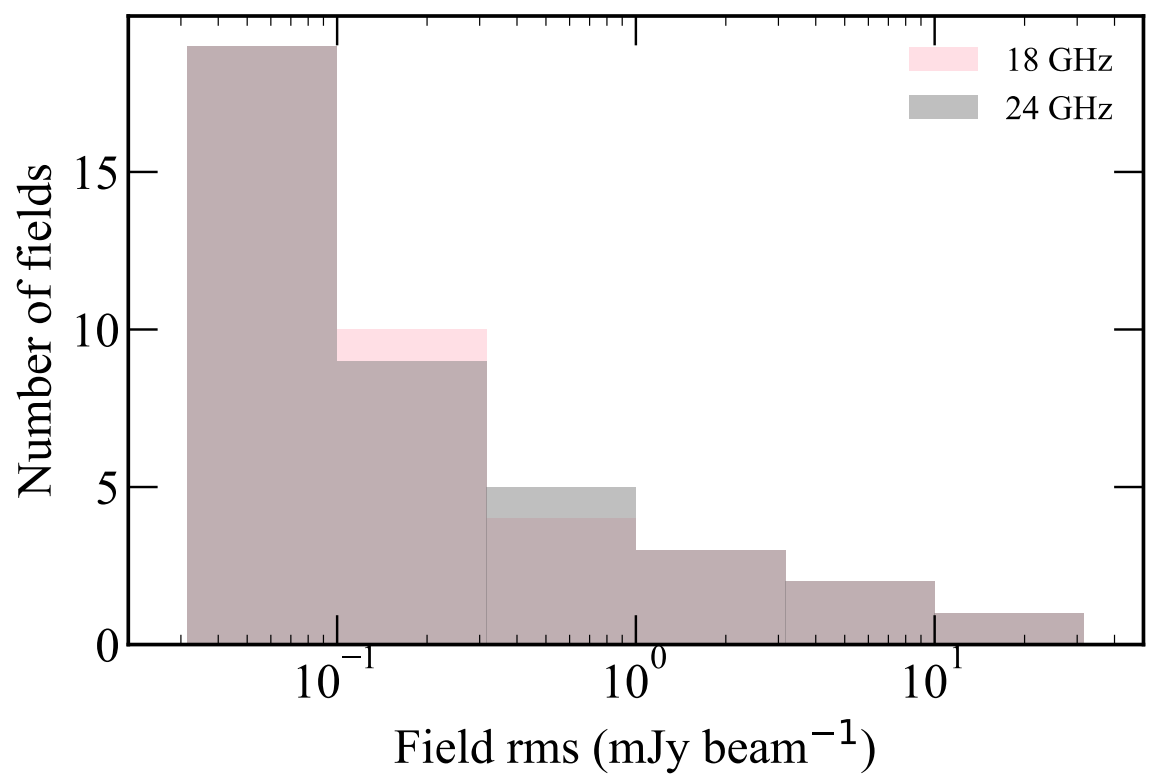

**Figure 1.** Histogram of the number of fields observed as a function of the map rms noise at both frequencies. The data have been binned using a value of 0.5 dex.

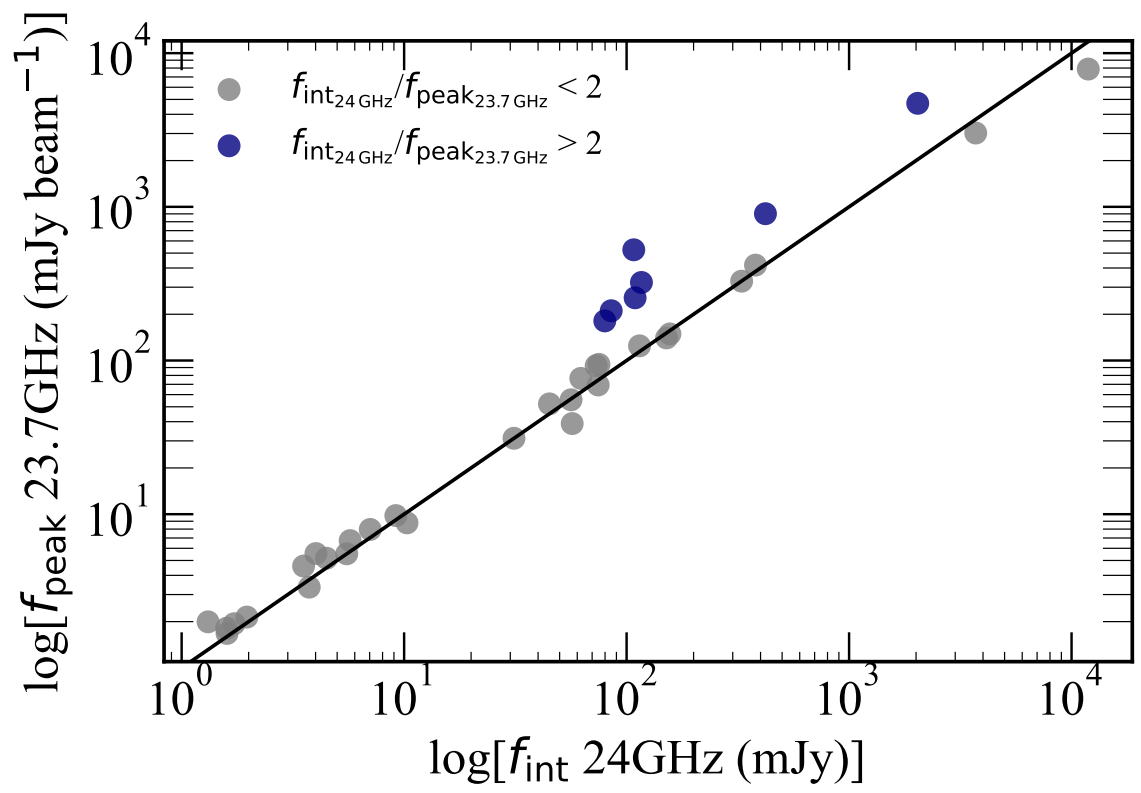

**Figure 2.** Distribution of the integrated flux density at 24 GHz from this study against the peak flux densities at 23.7 GHz in Paper I for the detected radio sources. Sources shown as blue circles are probably affected by spatial filtering. The solid black line represents the line of equality.

### 2.2 Data reduction and imaging

The calibration and reduction of these data were performed using the MIRIAD reduction package (Sault et al. 1995) following standard ATCA procedures. We performed a flagging procedure to eliminate any radio-frequency interference (RFI) and poor data. The procedure was performed iteratively, until all the RFI signals were removed from the visibility plots. The calibrated data were imaged and cleaned using the MIRIAD `CLEAN` algorithm using a robust weighting of 0.5 and we set a limit of 200 cleaning components or until the first negative component was encountered. All our fields converged at the first negative component. Additionally, we corrected the primary beam using the `LINMOS` function to generate a final map. This resulted in an image size of 50 × 50 arcsec using a pixel size of ∼0.2 arcsec. The final maps were made by combining the datasets for all four days to improve the *uv*-coverage and signal-to-noise ratio. In Figure 1, we present a histogram of the rms noise of each field. The rms noise levels were estimated from emission-free regions close to the centre of the reduced image. Median noise values are ∼0.13 mJy beam$^{-1}$, which is consistent with the theoretical noise values. Due to the nature of interferometric observations, the largest well imaged structures at 18 and 24 GHz are 5.5 and 4.0 arcsec, respectively. We present the observational parameters in Table 2.

**Table 1.** The CABB theoretical sensitivities for 18 GHz and 24 GHz, respectively. The synthesised beam is presented as beam major × beam minor. The noise values have been estimated assuming 10 minutes on-source integration time and stable weather conditions.

| Target Group | Average Declination (°) | Synthesised beam 18 GHz (arcsec) | 24 GHz (arcsec) | Theoretical RMS 18 GHz (mJy beam$^{-1}$) | 24 GHz (mJy beam$^{-1}$) |
|---|---|---|---|---|---|
| 1 | −52 | 0.61 × 0.48 | 0.46 × 0.36 | 0.094 | 0.160 |
| 2 | −46 | 0.67 × 0.48 | 0.50 × 0.36 | 0.094 | 0.160 |
| 3 | −25 | 1.14 × 0.48 | 0.86 × 0.36 | 0.098 | 0.166 |
| 4 | −15 | 2.66 × 0.48 | 1.41 × 0.36 | 0.100 | 0.168 |

### 2.3 Source parameters and final catalogue

The reduced maps were examined for compact, high surface brightness sources using a nominal 3$\sigma$ detection threshold, where $\sigma$ refers to the image rms noise level. The radio maps that had a peak flux density above the detection threshold were visually inspected to confirm the presence of the radio source and identify emission from over resolved sources and bright sidelobes.

In total, we have detected radio-continuum emission above 3 $\sigma$ in 35 of 39 observed fields, corresponding to a recovery rate of ∼90 per cent. The radio emission consists of a single compact radio source in 22 fields; a further 12 fields contain a single extended radio source, and the remaining field contains two distinct radio sources. We have identified seven fields (i.e., G002.614+0.134, G011.937−0.616, G014.104+0.092, G331.130−0.243, G332.826−0.549, G333.163−0.100, G340.056-0.244) which display evidence of large-scale, over-resolved emission (i.e., $f_{int_{24\,GHz}}/f_{peak_{23.7\,GHz}} > 2$). Each of these radio sources has a lower integrated flux density at 18 and 24 GHz when compared with the low-resolution 23.7-GHz peak flux density of Paper I suggesting a loss of flux in the high-resolution maps as a result of spatial filtering.

Figure 2 presents a comparison between the integrated flux densities at 24 GHz in this work and the peak flux densities at 23.7 GHz from Paper I. This plot shows that the majority of our radio sources cluster around the line of equality, demonstrating that there is no significant loss in flux density between the two spatial resolutions. The outliers (blue) correspond to the above seven over-resolved radio sources that are removed from the final catalogue.

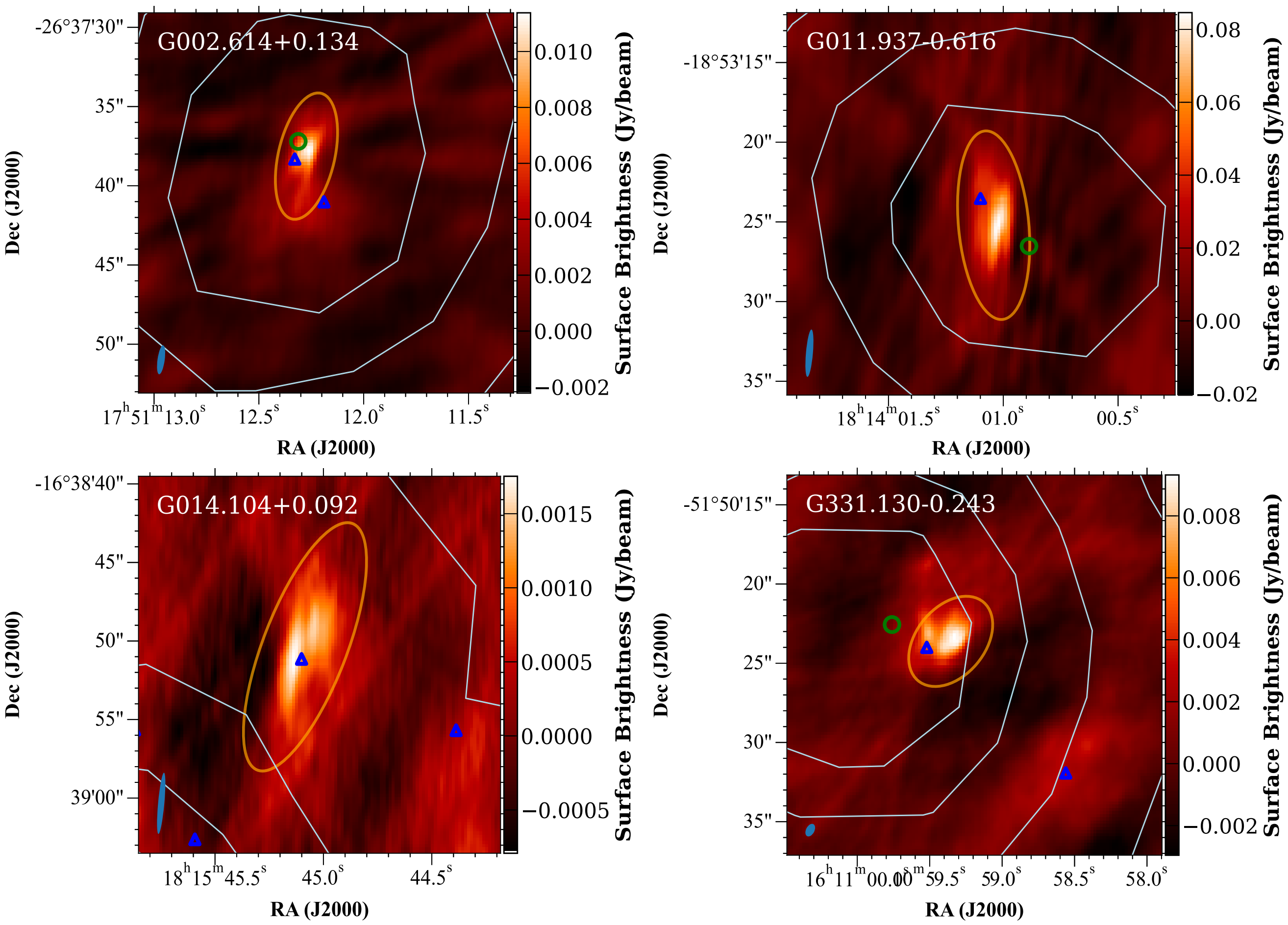

**Figure 3.** Examples of four radio fields that have been excluded. The remaining four fields can be found in Fig.A1 of the appendix. The orange ellipse shows the resultant fit to the radio emission while the green circles show the position of the methanol maser(s) located in the field. The blue triangles show the position of any low-frequency (5-GHz) radio counterparts. The grey contours trace the 870-μm dust emission from ATLASGAL (Schuller et al. 2009). The filled blue ellipse in the bottom left-hand corner of each image indicates the size and orientation of the synthesised beam.

**Table 2.** Observational parameters from all observed fields. The field names and pointing centres of the observations are based on the radio names and positions.

| Field name | RA | Dec. | Field rms | | Beam size 18 GHz | | | Beam size 24 GHz | | |
|---|---|---|---|---|---|---|---|---|---|---|
| | (J2000) | (J2000) | 18 GHz | 24 GHz | Major | Minor | Position angle | Major | Minor | Position angle |
| | (h:m:s) | (d:m:s) | (mJy) | (mJy) | (arcsec) | (arcsec) | (°) | (arcsec) | (arcsec) | (°) |
| G002.143+0.009 | 17:50:36.02 | −27:05:46.9 | 0.1 | 0.1 | 1.8 | 0.5 | −12.4 | 1.7 | 0.3 | −11.1 |
| G002.614+0.134 | 17:51:12.27 | −26:37:38.8 | 0.2 | 0.2 | 1.9 | 0.5 | −11.6 | 1.8 | 0.3 | −10.4 |
| G003.438−0.349 | 17:54:55.66 | −26:09:47.4 | 0.1 | 0.2 | 2.0 | 0.5 | −12.9 | 1.9 | 0.3 | −11.2 |
| G003.910+0.001 | 17:54:38.75 | −25:34:44.7 | 0.5 | 0.4 | 2.1 | 0.5 | −12.1 | 2.0 | 0.3 | −10.5 |
| G004.680+0.277 | 17:55:18.83 | −24:46:29.5 | 0.1 | 0.1 | 2.2 | 0.5 | −11.3 | 2.1 | 0.3 | −10.0 |
| G005.885−0.392 | 18:00:30.39 | −24:04:00.1 | 15.9 | 16.6 | 2.2 | 0.5 | −11.2 | 2.1 | 0.3 | −9.9 |
| G011.904−0.141 | 18:12:11.43 | −18:41:29.4 | 0.3 | 0.3 | 3.8 | 0.4 | −7.9 | 3.2 | 0.3 | −7.3 |
| G011.937−0.616 | 18:14:01.05 | −18:53:24.9 | 1.1 | 1.9 | 3.8 | 0.4 | −7.9 | 3.1 | 0.3 | −7.4 |
| G012.112−0.127 | 18:12:33.60 | −18:30:06.8 | 0.1 | 0.1 | 4.0 | 0.4 | −7.5 | 3.3 | 0.3 | −7.0 |

Note: only a small portion of the data is provided here, the full table is only available in electronic form.

We present maps of four of these sources in Figure 3. The strong correlation of fluxes for the compact radio sources also indicates that there is little variability in the brightness of these sources over the ten year period between the high- and low-resolution observations.

In Section 2.1, we mentioned that, for the late-rising sources during the observations, the synthesised beam is significantly elongated. We identify five fields that fall within this category. This can be seen visually as the sources appear elongated and as a result we are unable to constrain the sizes accurately. Of the five fields identified, one field contains two radio sources (G014.604+0.016). The fluxes extracted from this region are unlikely to be reliable and so the two sources in this map are excluded from further analysis. The source, G018.735−0.227, in the second field, shows the most significant elongation, and is reliably detected at 18 GHz (SNR = 9.5). However, at 24 GHz, we are unable to deconvolve the emission. In this case, we use the 18 GHz fit to estimate upper limits for the flux densities at 24 GHz. The remaining three fields are included, with a note indicating that the source sizes are less reliable.

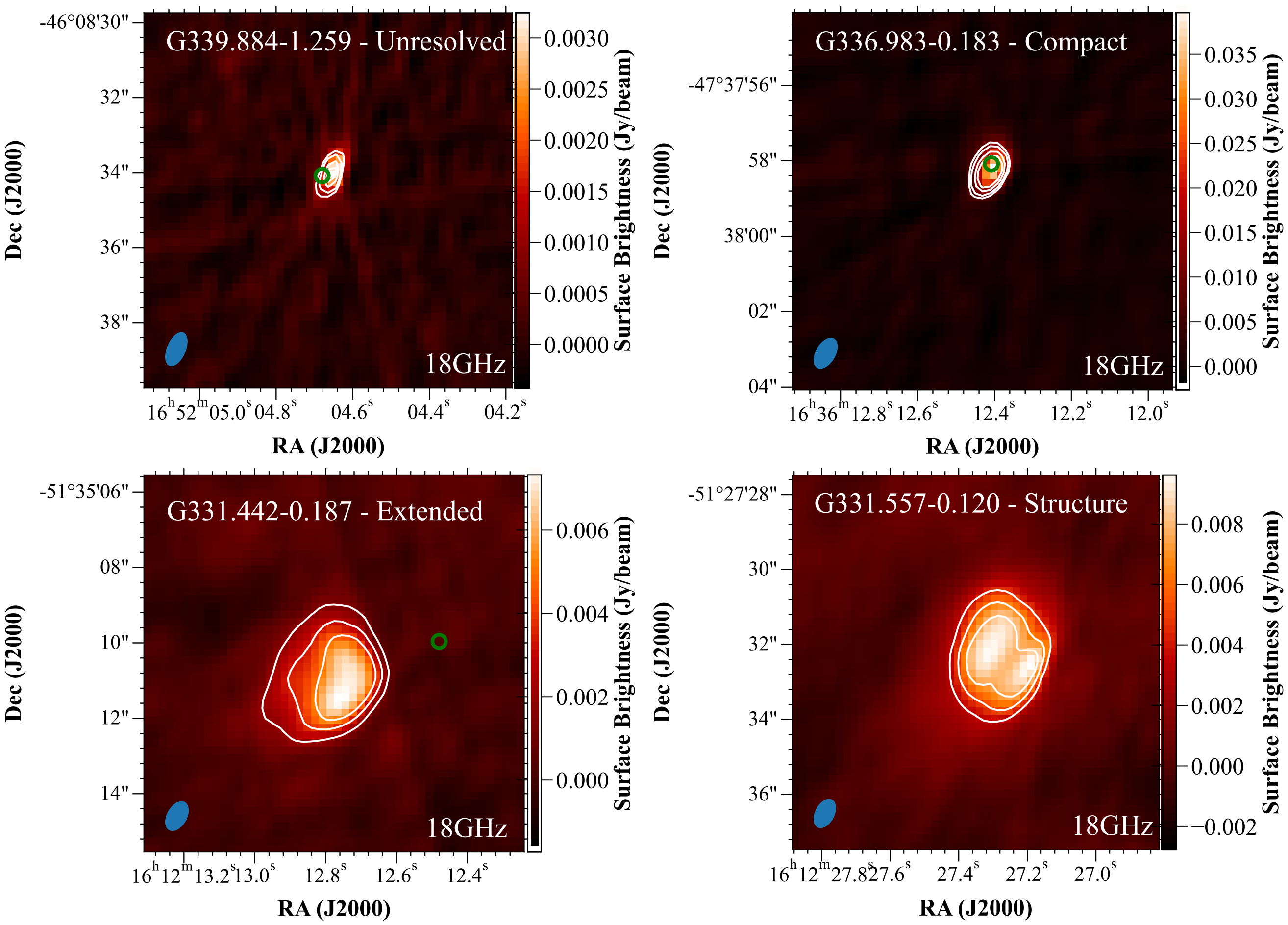

**Figure 4.** Examples of four radio sources with different morphologies. The top left panel presents an unresolved radio source (G339.884−1.259), the top right is compact radio source G336.983−0.183, the bottom left panel shows an extended source (G331.442−0.187) and the bottom right shows a structured source (G331.557−0.120). The white contours of each map start at a level of $3\sigma$ and are equally spaced by $3\sigma$. The orange ellipse shows the resultant fit from IMFIT to the radio emission while the green circles show the position of the methanol maser(s) located in the field. The filled blue ellipse in the bottom left-hand corner of each image indicates the size and orientation of the synthesised beam.

There are four non-detections (G012.112−0.127, G014.229−0.510, G331.710+0.604, G338.280+0.542) in this study and these objects were weak detections at 23.7 GHz of Paper I (i.e., SNR < 3) with average noise values of ∼0.5 mJy beam$^{-1}$. Our high-frequency data provides better sensitivity (i.e., ∼0.08 mJy beam$^{-1}$). It is possible that these objects were false detections at 23.7 GHz in Paper I as a result of higher noise values. Alternatively, given our limited sensitivity to spatial scales greater than 5 arcsec, it is possible that the emission from these objects could have been filtered out, resulting in them not being detected in these data. Another possibility is that they are variable HII regions as discussed in Yang et al. (2023).

We use the MIRIAD task IMFIT to determine the positions, peak and integrated flux densities and sizes of the remaining 27 compact radio detections. This was achieved by carefully drawing a polygon around the detection as an input to IMFIT where a two-component Gaussian is fitted around the emission.

## 3 RESULTS

### 3.1 Observational results

Our high-resolution catalogue is given in Table 3 and contains the extracted parameters for the 27 compact radio sources detected at 18 and 24 GHz. We present the radio name, RA, Dec, peak and integrated flux densities and FWHM major, minor axis sizes and position angle in Columns 1–16 of Table 3. In Column 15, we present the deconvolved major and minor source sizes, along with the position angle, determined from `IMFIT` using the highest resolution map where the source is detected. Additionally, in Column 16 we provide the geometric deconvolved source size. Five radio detections are unresolved at 18 GHz; these are indicated as dashes in Table 3.

We categorise our sample of 27 radio detections into four morphologies; unresolved, compact, extended and structured and in Figure 4 we present examples of each of these. The contouring highlights regions of low and high dynamic ranges, which enables us to differentiate between compact, bright radio emission and extended structures with low surface brightness. We have identified five sources that are unresolved at 18 GHz and have y-factors close to 1 (see the upper left panel of Fig. 4). A compact radio source

**Table 3:** Extracted source parameters from our detected radio sources. Column 1 is the radio name, which is a combination of the Galactic longitudes and latitudes determined by fitting ellipsoidal Gaussian using IMFIT, Columns 2 and 3 are the Right Ascension and Declination of each source, respectively. Columns 4−9 present 18-GHz source parameters and columns 10−16 present 24-GHz source parameters. Sources with − are those for which sizes could not be deconvolved. Sources with a † indicate those with a poor synthesised beam.

| | RA | Dec. | 18GHz | | | | | | 24GHz | | | | | | |
|---|---|---|---|---|---|---|---|---|---|---|---|---|---|---|---|
| Radio name (1) | J2000 (h:m:s) (2) | J2000 (d:m:s) (3) | $f_{peak}$ (mJy beam$^{-1}$) (4–5) | $\Delta f_{peak}$ | $f_{int}$ (mJy) (6–7) | $\Delta f_{int}$ | $\theta_{maj} \times \theta_{min}$ (PA) (arcsec × arcsec (°)) (8–9) | $\theta_{maj_d} \times \theta_{min_d}$ (PA$_d$) | $f_{peak}$ (mJy beam$^{-1}$) (10–11) | $\Delta f_{peak}$ | $f_{int}$ (mJy) (12–13) | $\Delta f_{int}$ | $\theta_{maj} \times \theta_{min}$ (PA) (arcsec × arcsec (°)) (14–15) | $\theta_{maj_d} \times \theta_{min_d}$ (PA$_d$) | $\theta_s$ (arcsec) (16) |
| G002.143+0.009† | 17:50:36.02 | −27:05:46.9 | 2.3 | 0.17 | 4.2 | 0.56 | 2.3 × 0.6 (−8.4) | 1.5 × 0.4 (−8.1) | 1.8 | 0.36 | 3.5 | 1.1 | 2.3 × 0.5 (−7.5) | 1.5 × 0.3 (−3.3) | 0.70 |
| G003.438−0.349† | 17:54:55.65 | −26:09:47.4 | 9.1 | 0.58 | 89 | 13 | 4.1 × 2.1 (+0.1) | 3.6 × 2.0 (+3.6) | 5.2 | 0.36 | 75 | 13 | 4.2 × 1.9 (−0.9) | 3.9 × 1.9 (+1.3) | 2.70 |
| G003.910+0.001† | 17:54:38.75 | −25:34:44.7 | 37 | 3.4 | 76 | 10 | 3.7 × 0.6 (−11) | 3.0 × 0.3 (−10) | 33 | 3.0 | 73 | 9.6 | 3.4 × 0.4 (−9.7) | 2.7 × 0.3 (−9.2) | 0.87 |
| G004.680+0.277† | 17:55:18.83 | −24:46:29.5 | 6.5 | 0.64 | 9.8 | 1.4 | 3.3 × 0.5 (−9.8) | – | 5.5 | 0.61 | 9.2 | 1.5 | 3.0 × 0.4 (−9.5) | 2.2 × 0.2 (−9.1) | 0.65 |
| G005.885−0.392† | 18:00:30.38 | −24:04:00.5 | 470 | 49 | 15000 | 9200 | 6.8 × 4.5 (+57) | 6.8 × 4.1 (+60) | 280 | 32 | 12000 | 5400 | 6.5 × 4.0 (+31) | 6.4 × 3.8 (+35) | 4.90 |
| G011.904−0.141† | 18:12:11.43 | −18:41:29.6 | 15 | 2.2 | 41 | 12 | 4.6 × 0.9 (−1.2) | 3.5 × 0.7 (+1.5) | 14 | 1.6 | 40 | 8.7 | 3.5 × 0.7 (−3.5) | 2.4 × 0.6 (−1.9) | 1.20 |
| G012.199−0.034† | 18:12:23.60 | −18:22:53.5 | 6.4 | 0.49 | 54 | 8.2 | 5.7 × 2.1 (−1.6) | 4.7 × 2.0 (+0.1) | 3.5 | 0.24 | 57 | 14 | 7.2 × 1.9 (+2.8) | 6.6 × 1.9 (+4.2) | 3.50 |
| G012.208−0.102† | 18:12:39.69 | −18:24:20.7 | 22 | 1.0 | 160 | 13 | 5.2 × 2.0 (−11) | 4.1 × 1.9 (−16) | 12 | 0.64 | 150 | 24 | 6.3 × 1.7 (−11) | 5.7 × 1.7 (−12) | 3.10 |
| G018.461−0.004† | 18:24:36.34 | −12:51:05.3 | 64 | 5.1 | 390 | 70 | 7.5 × 1.6 (−3.0) | 5.9 × 1.6 (−2.1) | 37 | 4.2 | 380 | 120 | 10 × 1.3 (0.18) | 9.3 × 1.2 (+1.3) | 3.40 |
| G018.665+0.029† | 18:24:52.60 | −12:39:20.0 | 2.2 | 0.18 | 5.9 | 0.86 | 6.1 × 0.9 (−2.7) | 3.7 × 0.8 (+0.1) | 1.5 | 0.16 | 5.5 | 1.0 | 6.5 × 0.8 (−3.1) | 4.3 × 0.7 (−1.8) | 1.80 |
| G018.735−0.227† | 18:25:56.47 | −12:42:49.0 | 0.62 | 0.15 | 1.7 | 1.3 | 7.9 × 0.7 (−2.0) | 6.2 × 0.5 (−0.6) | < 0.54 | – | < 1.4 | – | – | – | – |
| G326.475+0.703 | 15:43:16.62 | −54:07:14.9 | 0.75 | 0.053 | 0.84 | 0.19 | 0.8 × 0.6 (−33) | – | 1.2 | 0.36 | 2.0 | 1.1 | 0.8 × 0.4 (−25) | 0.7 × 0.1 (−24) | 0.27 |
| G327.402+0.445 | 15:49:19.32 | −53:45:14.4 | 41 | 1.6 | 120 | 6.9 | 1.6 × 0.9 (−17) | 1.1 × 0.4 (−11) | 30 | 2.3 | 110 | 13 | 1.4 × 0.7 (−13) | 1.1 × 0.6 (−9.8) | 0.80 |
| G331.442−0.187 | 16:12:12.75 | −51:35:10.9 | 7.2 | 0.34 | 78 | 9.9 | 2.5 × 2.0 (−23) | 2.0 × 1.1 (−21) | 3.8 | 0.2 | 75 | 8.2 | 2.7 × 1.8 (−29) | 2.1 × 1.3 (−30) | 1.70 |
| G331.542−0.067 | 16:12:09.02 | −51:25:47.5 | 130 | 5.1 | 320 | 18 | 1.3 × 0.9 (−37) | 1.0 × 0.7 (−38) | 80 | 4.1 | 330 | 25 | 1.3 × 0.8 (−25) | 1.2 × 0.7 (−25) | 0.91 |
| G331.557−0.120 | 16:12:27.28 | −51:27:32.4 | 11 | 0.33 | 170 | 7.4 | 3.0 × 2.3 (−25) | 1.7 × 1.3 (−24) | 6.1 | 0.29 | 160 | 13 | 3.1 × 2.2 (−19) | 1.8 × 1.4 (−18) | 1.60 |
| G332.987−0.487 | 16:20:37.77 | −50:43:49.6 | 6.4 | 0.78 | 9.2 | 1.7 | 1.1 × 0.6 (−15) | – | 7.0 | 1.1 | 12 | 2.9 | 0.9 × 0.5 (−9.2) | 0.7 × 0.2 (+3.6) | 0.36 |
| G333.029−0.063 | 16:18:56.86 | −50:23:54.2 | 15 | 1.6 | 25 | 3.7 | 1.2 × 0.6 (−22) | 0.9 × 0.1 (−11) | 13 | 1.7 | 31 | 6.7 | 1.2 × 0.6 (−11) | 1.0 × 0.3 (−5.6) | 0.59 |
| G333.135−0.432 | 16:21:02.95 | −50:35:12.6 | 360 | 25 | 3400 | 860 | 2.4 × 1.9 (+38) | 2.4 × 1.7 (+40) | 230 | 17 | 3700 | 770 | 2.8 × 1.6 (+6.0) | 2.7 × 1.5 (+7.7) | 2.05 |
| G336.983−0.183 | 16:36:12.41 | −47:37:58.3 | 39 | 1.5 | 52 | 3.0 | 0.8 × 0.3 (−25) | 0.6 × 0.4 (−8.2) | 33 | 2.4 | 56 | 6.0 | 1.0 × 0.5 (−16) | 0.7 × 0.4 (−12) | 0.52 |
| G338.566+0.110 | 16:41:07.15 | −46:15:30.4 | 1.6 | 0.091 | 9.7 | 0.91 | 2.1 × 1.5 (−16) | 1.9 × 1.4 (−13) | 9.8 | 0.081 | 10 | 1.5 | 2.4 × 1.3 (−7.9) | 2.3 × 1.2 (−6.6) | 1.71 |
| G338.925+0.556 | 16:40:33.54 | −45:41:37.6 | 1.1 | 0.15 | 1.4 | 0.33 | 1.3 × 0.6 (−15) | – | 1.2 | 0.037 | 1.6 | 0.14 | 1.0 × 0.4 (−19) | 0.5 × 0.2 (−22) | 0.27 |
| G339.282+0.136 | 16:43:43.08 | −45:42:08.6 | 2.6 | 0.26 | 3.1 | 0.57 | 1.2 × 0.6 (−23) | 0.7 × 0.3 (−24) | 2.0 | 0.36 | 3.7 | 1.2 | 1.0 × 0.6 (−21) | 0.6 × 0.4 (−31) | 0.50 |
| G339.622−0.121 | 16:46:06.00 | −45:36:44.0 | 1.0 | 0.16 | 1.7 | 0.69 | 1.3 × 0.7 (−14) | 0.9 × 0.4 (−5.5) | 0.95 | 0.29 | 1.7 | 1.2 | 1.1 × 0.5 (−12) | 0.8 × 0.3 (−7.4) | 0.47 |
| G339.681−1.207 | 16:51:06.22 | −46:15:58.1 | 0.92 | 0.15 | 1.2 | 0.36 | 1.2 × 0.6 (−20) | 0.9 × 0.2 (−16) | 1.4 | 0.39 | 1.3 | 0.81 | 0.7 × 0.3 (−11) | 0.8 × 0.2 (−26) | 0.36 |
| G339.884−1.259 | 16:52:04.65 | −46:08:34.0 | 3.4 | 0.34 | 3.7 | 0.78 | 0.8 × 0.4 (−18) | – | 4.3 | 0.54 | 5.7 | 1.3 | 0.9 × 0.4 (−12) | 0.5 × 0.1 (+4.1) | 0.26 |
| G339.980−0.539 | 16:49:14.72 | −45:36:32.4 | 44 | 2.5 | 110 | 9.0 | 1.5 × 0.9 (−14) | 1.1 × 0.7 (−5.7) | 28 | 1.1 | 100 | 6.0 | 1.4 × 0.7 (−9.4) | 1.2 × 0.6 (−5.2) | 0.86 |

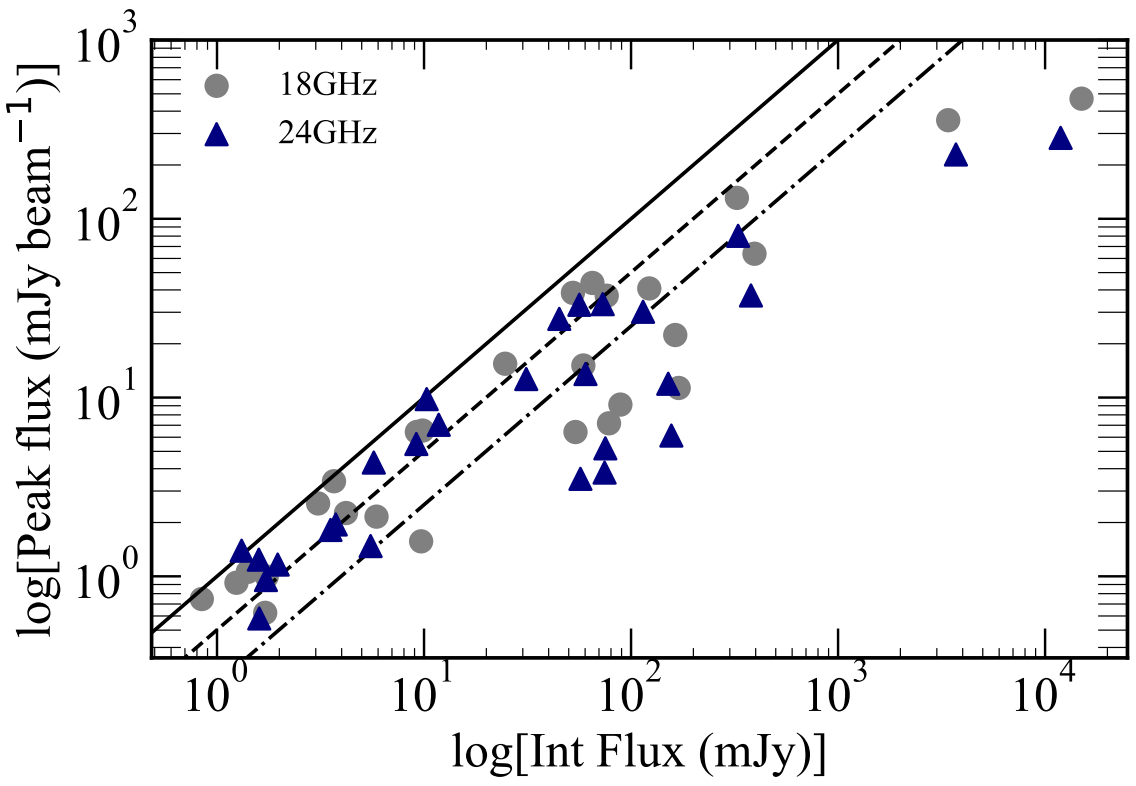

**Figure 5.** Distribution of the integrated flux density vs peak flux density for the 27 radio detections at 18 and 24 GHz. The solid line represents a line of equality where the flux densities are the same, indicating the radio source is unresolved. The dashed line represents a line with $y = 2$ suggesting the radio source is compact. The dash-dotted line has a y−factor of 4; sources with y−factors above this are considered to be extended.

is described to have a y−factor < 2 and shows visual signs of extension but no structure (see lower-left panel of Fig. 4). A further eleven sources are identified as being extended. We find three radio sources that have complex sub-structure and an example is shown in the lower-right panel of Fig. 4.

In Figure 5, we present a scatter plot of the y−factor at 18 and 24 GHz for the 27 radio detections. Inspection of the plot reveals that the distribution is in agreement with the morphologies characterised above. In total, we have identified 14 radio sources that have a y−factor ≤ 2 that we have classified to be compact, and 13 radio sources that are extended.

### 3.2 Radio source properties and SED models

In this section, we investigate the physical characteristics of the 27 radio detections, such as sizes, emission measure, electron densities, Lyman-continuum flux and turnover frequency. These can be estimated from the observational angular sizes and flux densities at a given frequency, assuming the continuum emission comes from a homogeneous, optically thin ionized gas (Urquhart et al. 2013a; Kalcheva et al. 2018; Yang et al. 2021).

The physical properties of young HII regions can be underestimated or overestimated when using the observational parameters from a single frequency. This is the case primarily for two reasons: first, the young HII region may be optically thick at the observing frequency and second, the angular size of the object is dependant on the observing frequency (Yang et al. 2019). Therefore, in order to accurately estimate the physical properties of young HII regions, it is essential to know their radio spectral energy distribution (SED) over a wide frequency range that cover both optically thick and thin regimes.

Using the available multi-wavelength data from CORNISH-North 5 GHz (Purcell et al. 2013), CORNISH-South 5 GHz (Irabor et al. 2023) and MAGPIS 5 GHz Becker et al. 1994), we construct radio SEDs for the 27 HII regions. The $1\sigma$ sensitivity for CORNISH-North, South and MAGPIS is 0.35, 0.11 and 0.179 mJy beam$^{-1}$. In cases where fluxes from different surveys overlap for our radio sources, we prioritise using the 5-GHz flux densities from the CORNISH-North and South surveys due to their greater consistency within the context of this study. We find that 21 of the 27 radio sources are associated with unresolved or compact 5 GHz radio counterparts and are broken down as: 5 MAGPIS, 6 CORNISH-North and 10 CORNISH-South. For the remainder 6 sources that are not detected at 5 GHz we use the $3\sigma$ rms noise from the 5 GHz maps to determine a lower limit for the flux.

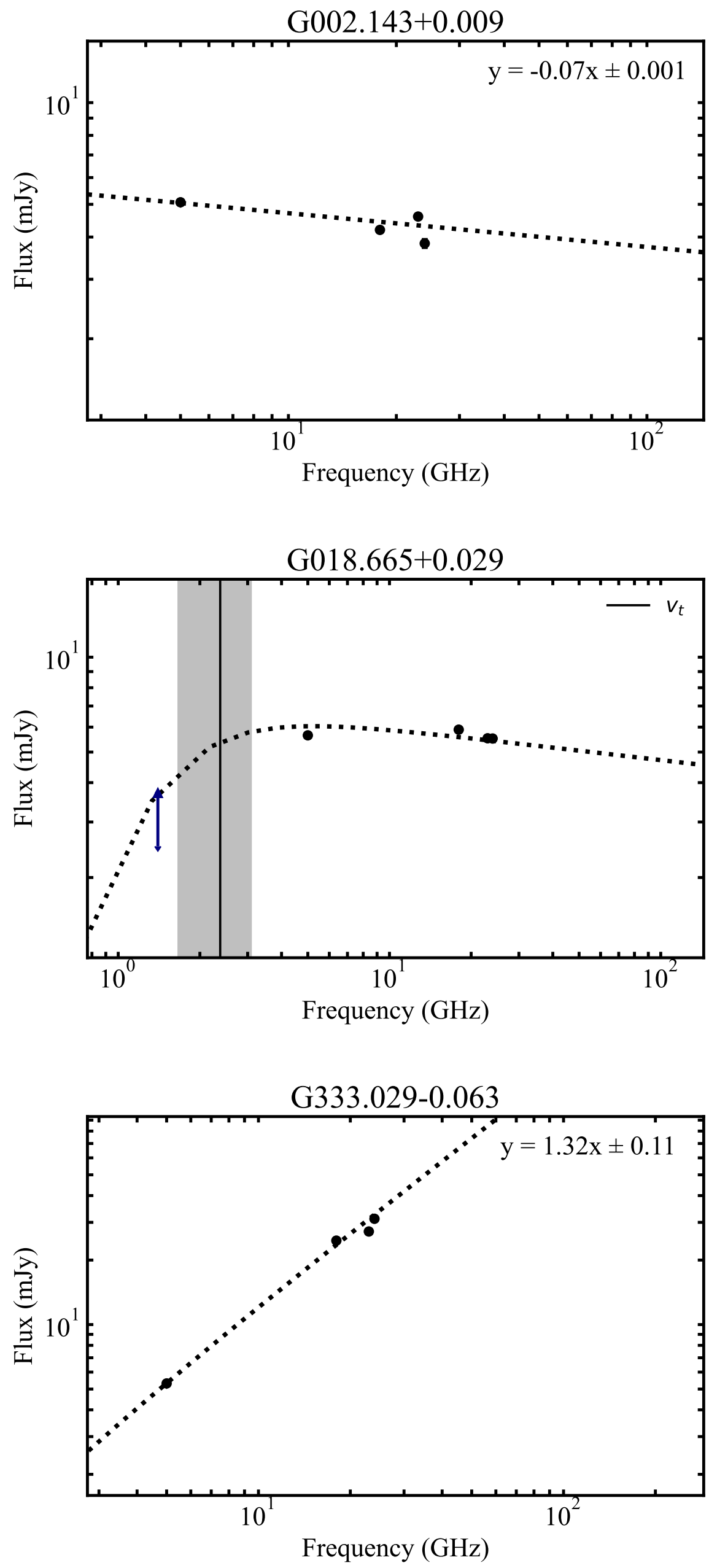

**Figure 6.** Three examples of radio SED fitting for varying optical depth. Upper panel: SED fitting to the integrated flux density for an optically thin source (G002.143+0.009). Middle panel: SED fitting for radio source G018.665+0.029, an example of a source constrained by both optically thick and thin regimes (Yang et al. 2021). The vertical solid line represents the turnover frequency ($\nu_t$) and the shaded region provides an estimation of the uncertainty. The blue triangle represents the 1.4-GHz peak flux density from THOR. Lower panel: SED fitting to the integrated flux density for an optically thick source (G333.029−0.063).

Each radio SED model is constructed using the standard uniform electron density model for a bound ionised HII region given by Mezger & Henderson (1967). The integrated flux density at a given frequency $\nu$ is given by Eqn. 1 using the Rayleigh Jeans approximation:

$$S_\nu = \frac{2k\nu^2\Omega T_e(1-e^{-\tau})}{c^2}, \quad (1)$$

where $k$ is the Boltzmann constant, $\Omega$ is the solid angle, $T_e$ is the electron temperature and $c$ is the speed of light. The optical depth $\tau$ of free-free radiation can be represented as a function of frequency (Mezger & Henderson 1967; Dyson & Williams 1997):

$$\tau \propto T_e^{-1.35}\nu^{-2.1}EM, \quad (2)$$

where we assume the electron temperature $T_e = 10^4$K (Dyson & Williams 1997) and $EM$ is the emission measure. The standard model of a radio SED is therefore, expected to have a rising spectrum at low frequencies $S_\nu \propto \nu^{+2}$ and a flat spectrum at high frequencies $S_\nu \propto \nu^{-0.1}$. Using the model provided by Yang et al. (2021) the radio SED for each source has two free parameters: the electron density ($n_e$) and the diameter (diam). We obtain the best estimate for the two parameters by fitting the available radio continuum over a range of frequencies. The uncertainties of the integrated flux densities are given as an output from IMFIT (see Table 3) and are taken into consideration during the fitting process. The emission measure can be calculated using the EM = ${n_e}^2$ × diam as demonstrated by Yang et al. (2021). The model assumes a spherical morphology and represents averaged estimates for the properties (e.g., $n_e$ and diameter) of optically thin ionized gas between 5 and 24 GHz.

In Figure 6 we present three examples of SEDs with varying optical depths. A complete set of SEDs for the 27 compact sources is provided in Figure B1 of the appendix, which can be found as online supplementary material. We have identified eleven radio SEDs that lack constraints spanning both optically thick and thin regimes. However, for five of these radio sources located between $l$ = 12°– 18°, we utilise the peak flux densities from THOR 1.4-GHz (Wang et al. 2018) and MAGPIS 1.4-GHz (White et al. 2005) data to constrain the low-frequency end of the SED and determine the turnover frequency. The additional 1.4-GHz fluxes are presented as upper limits to the peak flux density, as the resolutions of these surveys are significantly larger than the data presented here (~20 arcsec). The remaining six radio sources that lack constraints across both optically thick and thin regimes have been fitted with a linear function and their sizes have been taken from IMFIT. For these six sources, we use Equations 3-5 outlined in Section 3.3 to derive the electron density, emission measure, Lyman continuum and turnover frequencies. Among these six sources, two are optically thin between 5–24 GHz, whereas four remain optically thick at 24 GHz. Considering the optically thick nature of these four sources, it is likely that the parameters are underestimated and therefore, we present their physical parameters as upper limits to the size and lower limits to the electron density ($n_e$), emission measure and Lyman continuum flux ($N_{Ly}$).

### 3.3 Physical parameters of HII regions

Assuming that the emission of the detected radio continuum originates from homogeneous optically thin HII regions, we have calculated the physical parameters for 27 radio sources using the equations found in Mezger & Henderson (1967) and Rubin (1968). The kinematic distances from the literature are only available for 26 sources, therefore, our analysis will focus exclusively on these. We compare the results from the 21 optically thin, SED-derived parameters against the output from the equations in Fig. 8. Inspection of this figure reveals two outliers within our sample, these sources correspond to the two most elongated radio sources in our sample and the measured diameters are considered to be upper limits (these are shown as red upper limits). Excluding these two sources we find the sizes determined from the two methods are in very good agreement.

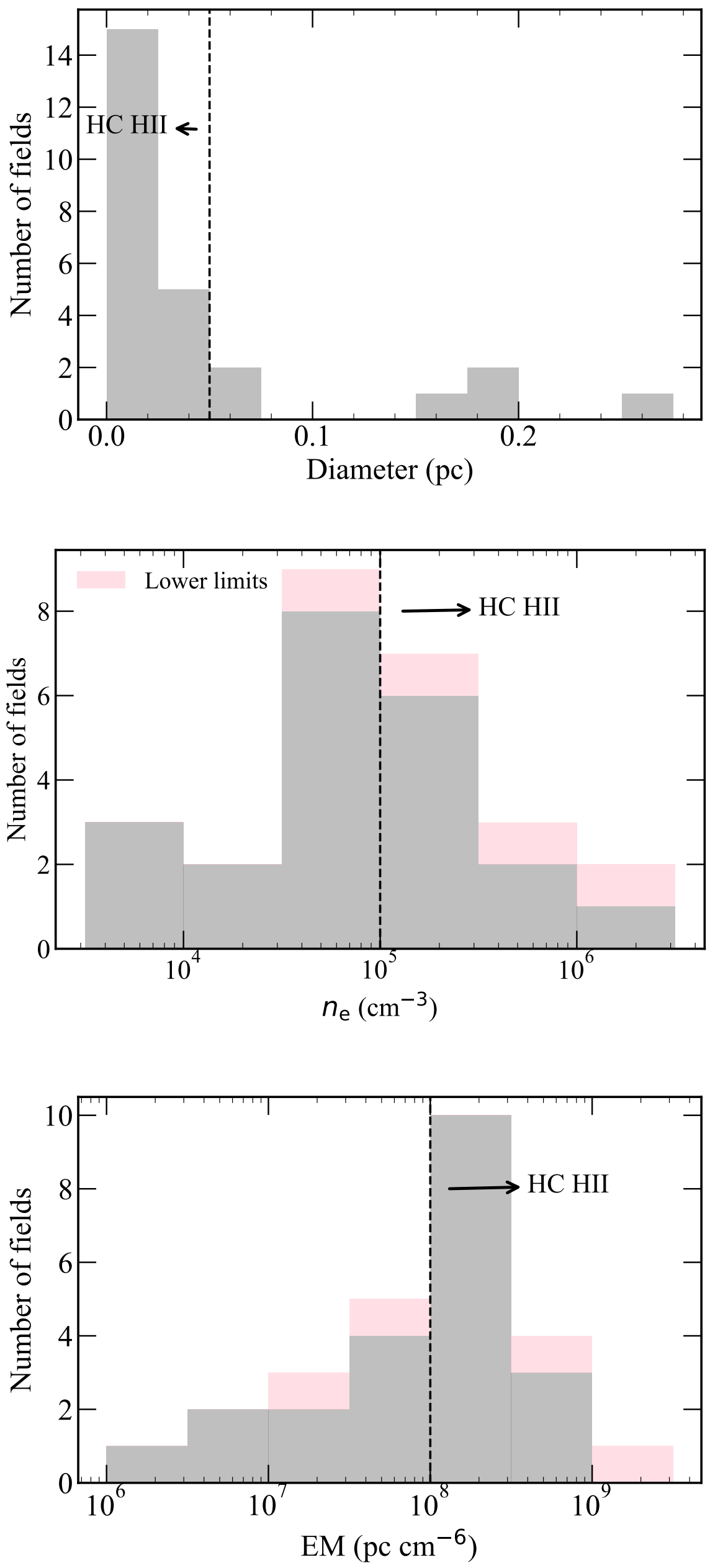

**Figure 7.** Histograms presenting the diameter (upper panel), electron density ($n_e$; middle panel) emission measure (lower panel) and of the derived physical parameters for the 26 HII regions. The bin sizes are 0.025 pc, 0.5 dex and 0.5 dex for diameter and $n_e$ and emission measure, respectively. The pink fraction of the distribution represents values that are lower limits. The black arrows emphasise the region of parameter space in which we expect to find HC HII regions.

**Table 4.** Derived physical parameters for 26 H II regions. The sources whose properties were not extracted via the SED fitting are denoted by a †.

| Radio name | Diameter (pc) | $n_e$ ($10^4$ cm$^{-3}$) | $EM$ ($10^7$ pc cm$^{-6}$) | $\log_{10}[N_{Ly}]$ (photons s$^{-1}$) | $\nu_t$ GHz | Spectral type | Classification |
|---|---|---|---|---|---|---|---|
| G003.438−0.349 | 0.2649 | 0.61 | 0.66 | 48.62 | – | O7 | UC H II |
| G003.910+0.001† | 0.0105 | 9.92 | 10.37 | 46.89 | 5.33 | B0.5 | HC H II |
| G004.680+0.277 | 0.0030 | 22.97 | 15.98 | 45.98 | 6.55 | B1 | HC H II |
| G005.885−0.392 | 0.0577 | 8.93 | 46.03 | 49.12 | 10.84 | O6 | UC H II |
| G011.904−0.141 | 0.0134 | 10.86 | 5.19 | 45.67 | 3.84 | B1 | HC H II |
| G012.199−0.034† | 0.1749 | 0.51 | 0.29 | 47.90 | 2.48 | O9.5 | UC H II |
| G012.208−0.102† | 0.1883 | 0.92 | 1.05 | 48.52 | 2.19 | O7.5 | UC H II |
| G018.461−0.004 | 0.1903 | 1.12 | 2.39 | 48.82 | 2.34 | O6.5 | UC H II |
| G018.665+0.029† | 0.0321 | 1.72 | 0.95 | 46.84 | 2.38 | B0.5 | HC H II– UC H II |
| G018.735−0.227 | 0.0046 | 21.77 | 21.90 | 46.49 | 7.61 | B0.5 | HC H II |
| G326.475+0.703† | 0.0032 | 7.16 | 1.82 | 45.53 | – | B1 | HC H II |
| G327.402+0.445 | 0.0258 | 5.42 | 7.59 | 47.51 | 4.60 | B0 | HC H II– UC H II |
| G331.442−0.187 | 0.0246 | 4.68 | 5.40 | 47.38 | 3.91 | B0 | HC H II– UC H II |
| G331.542−0.067 | 0.0308 | 7.29 | 16.40 | 48.03 | 6.64 | O6.5 | HC H II– UC H II |
| G331.557−0.120 | 0.0525 | 3.27 | 5.61 | 48.01 | 3.98 | O9.5 | UC H II |
| G332.987−0.487 | 0.0014 | 118.90 | 195.35 | 46.25 | – | B0.5 | HC H II |
| G333.029−0.063 | 0.0026 | 50.94 | 68.60 | 46.47 | – | B0.5 | HC H II |
| G333.135−0.432 | 0.0486 | 7.81 | 29.63 | 48.70 | 8.79 | O7 | HC H II– UC H II |
| G336.983−0.183 | 0.0099 | 14.98 | 22.16 | 47.17 | 7.66 | B0.5 | HC H II |
| G338.566+0.110 | 0.0048 | 19.93 | 18.87 | 46.48 | 7.09 | B0.5 | HC H II |
| G338.925+0.556 | 0.0008 | 109.97 | 95.42 | 46.54 | 15.34 | B0.5 | HC H II |
| G339.282+0.136 | 0.0025 | 29.29 | 21.27 | 45.99 | 8.51 | B1 | HC H II |
| G339.622−0.121 | 0.0007 | 88.01 | 54.26 | 45.17 | 11.73 | B1 | HC H II |
| G339.681−1.207 | 0.0008 | 53.11 | 21.30 | 45.93 | 7.51 | B1 | HC H II |
| G339.884−1.259† | 0.0025 | 17.91 | 5.31 | 45.58 | – | B1 | HC H II |
| G339.980−0.539 | 0.0425 | 4.90 | 10.21 | 48.09 | 5.30 | O9 | HC H II– UC H II |

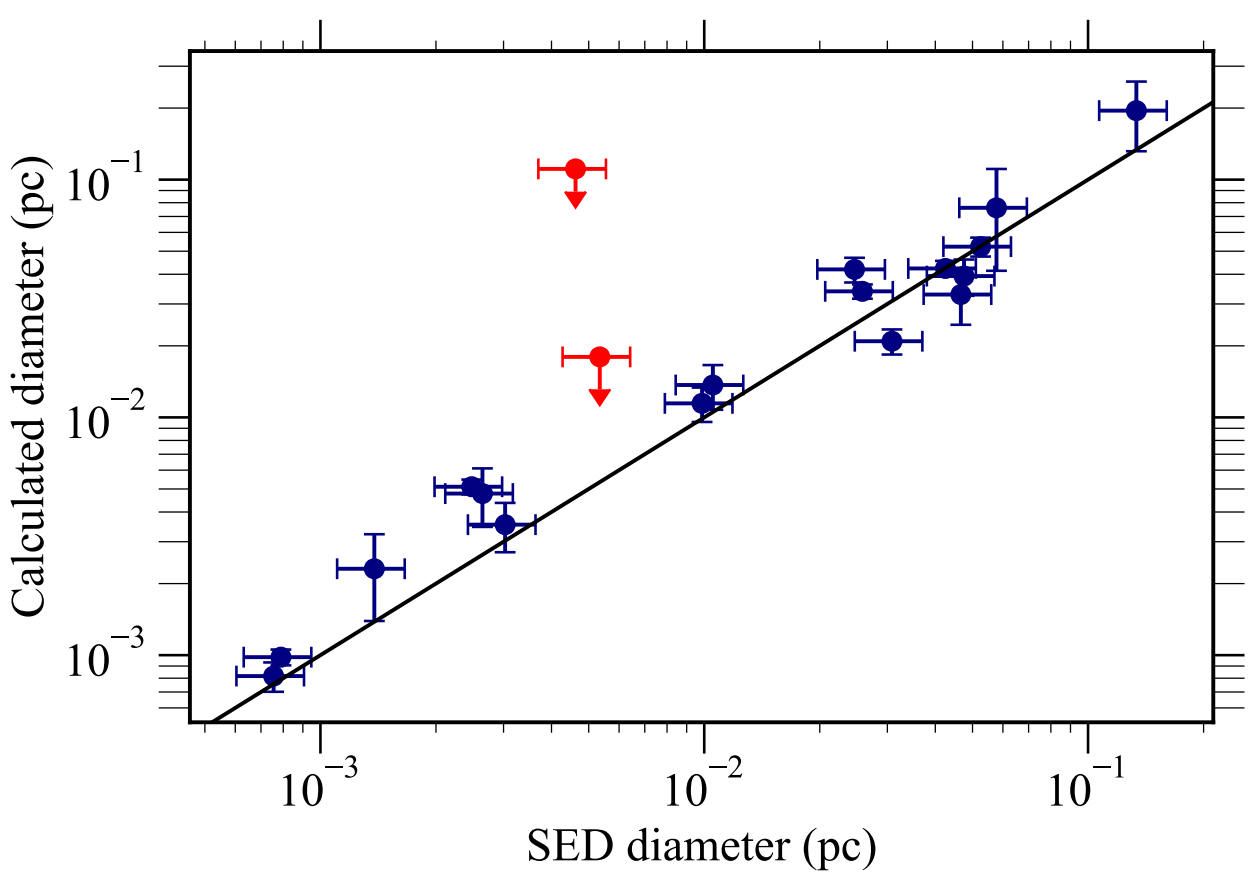

**Figure 8.** Distribution of the calculated sizes as a function of the SED derived sizes for the optically thin H II regions in this work. The red upper limits correspond to the two most elongated radio sources in our sample. The black solid line represents the line of equality.

To check the consistency of the parameters obtained from the SED model and equations we perform a chi-squared test where our null hypothesis states that the equation-calculated parameters are consistent with the values obtained from the SED fitting. Excluding the two elongated sources from the analysis, we find that the diameter, emission measure and electron density are comparable within the respective uncertainties ($p$-values > 0.7). We utilize the observational parameters extracted from the 24-GHz maps to constrain the physical parameters accurately. The linear diameter of the H II region was calculated using the deconvolved sizes listed in Table 3. We present the fitted SED parameters and derived physical properties in Table 4.

The emission measure ($EM$) and electron density ($n_e$) are calculated using Mezger & Henderson (1967); Rubin (1968):

$$\left(\frac{EM}{\text{cm}^{-6}\,\text{pc}}\right) = 1.7\times10^{7}\left(\frac{S_\nu}{\text{Jy}}\right)\left(\frac{\nu}{\text{GHz}}\right)^{0.1}\left(\frac{T_e}{10^4\text{K}}\right)^{0.35}\left(\frac{\theta}{\text{arcsec}}\right)^{-2},$$

$$\left(\frac{n_e}{\text{cm}^{-3}}\right) = 2.3\times10^{6}\left(\frac{S_\nu}{\text{Jy}}\right)^{0.5}\left(\frac{\nu}{\text{GHz}}\right)^{0.05}\left(\frac{T_e}{10^4\text{K}}\right)^{0.175} \times\left(\frac{d}{\text{pc}}\right)^{-0.5}\left(\frac{\theta}{\text{arcsec}}\right)^{-1.5}, \quad (3)$$

where $\nu$ is 24 GHz, $S_\nu$ is the integrated flux density at 24 GHz, $T_e$ is electron temperature, which is assumed to be $10^4$K, $\theta$ is the angular diameter and $d$ is the kinematic distance to the source, which has been taken from the literature (i.e., Urquhart et al. 2018, 2022). We do not have an accurate distance for one clump (AGAL002.142+00.009[2]) and so we are unable to calculate reliable measurements for the diameter and emission measure.

In Figure 7 we show the distribution of the diameter, electron density and emission measure using the values presented in Table 4. As previously mentioned, upper limits to the diameter and lower limits for the election density and emission measure were determined for the four optically thick sources (i.e., pink regions of the histogram). The vertical dotted line represents the typical criteria of HC H II regions. These plots confirm that our sample have parameter values similar to those of known HC Hɪɪ regions. The mean values for diameter, electron density and emission measure are given in Table 5.

[2] This source is located toward the Galactic centre where kinematic distances are unreliable.

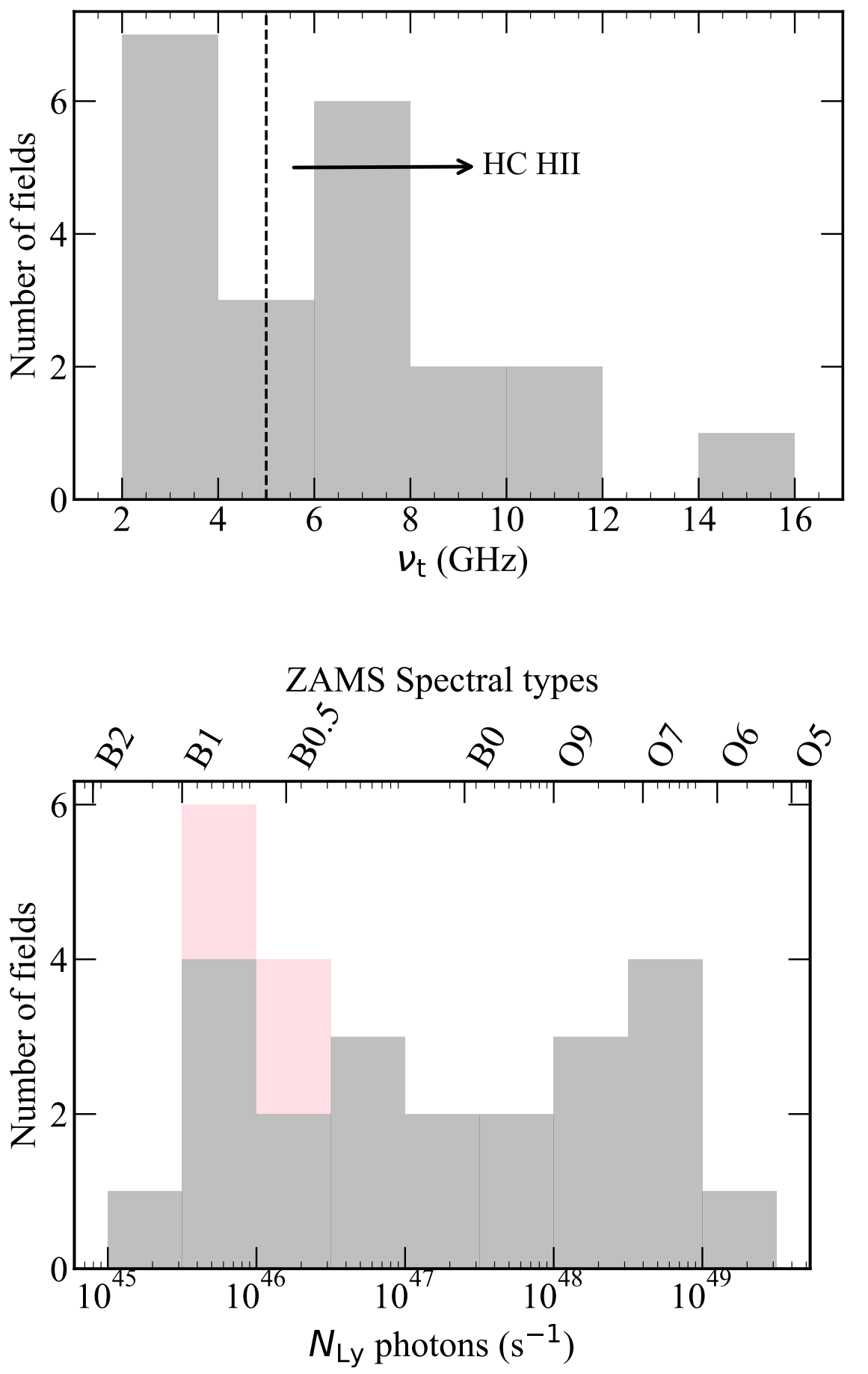

**Figure 9.** Histograms presenting the turnover frequency ($\nu_t$) and $N_{Ly}$ flux of the derived physical parameters for the 26 Hɪɪ regions. The bin sizes are 2 GHz and 0.5 dex for $\nu_t$ and $N_{Ly}$, respectively. The pink fraction of the distribution represents values that are lower limits. The black arrow in the upper panel emphasises the region of parameter space in which we expect to find HC Hɪɪ regions.

**Table 5.** Mean values of the physical parameters for HC Hɪɪ regions, UC Hɪɪ regions and intermediate objects identified in this work.

| Parameters | Size (pc) | $n_e$ ($cm^{-3}$) | EM ($pc\,cm^{-6}$) | $\nu_t$ (GHz) |
|---|---|---|---|---|
| HC Hɪɪ region | 0.007 | $10^{5.9}$ | $10^{8.7}$ | 10.4 |
| Intermediate objects | 0.03 | $10^{5.3}$ | $10^{8}$ | 7.9 |
| UC Hɪɪ region | 0.15 | $10^{4.2}$ | $10^{7.9}$ | 6.2 |

### 3.3.1 Turnover Frequency

The behaviour observed in the radio spectrum of an Hɪɪ region can vary depending on its optical depth, which is frequency dependant. The frequency for which the optical depth $\tau_\nu = 1$, is known as the turnover frequency ($\nu_t$ Kurtz & Hofner 2005). Using the formula provided by Wilson et al. (2013) assuming a homogeneous, optically thin Hɪɪ region, the turnover frequency can be expressed as a function of the emission measure and electron temperature $T_e$.

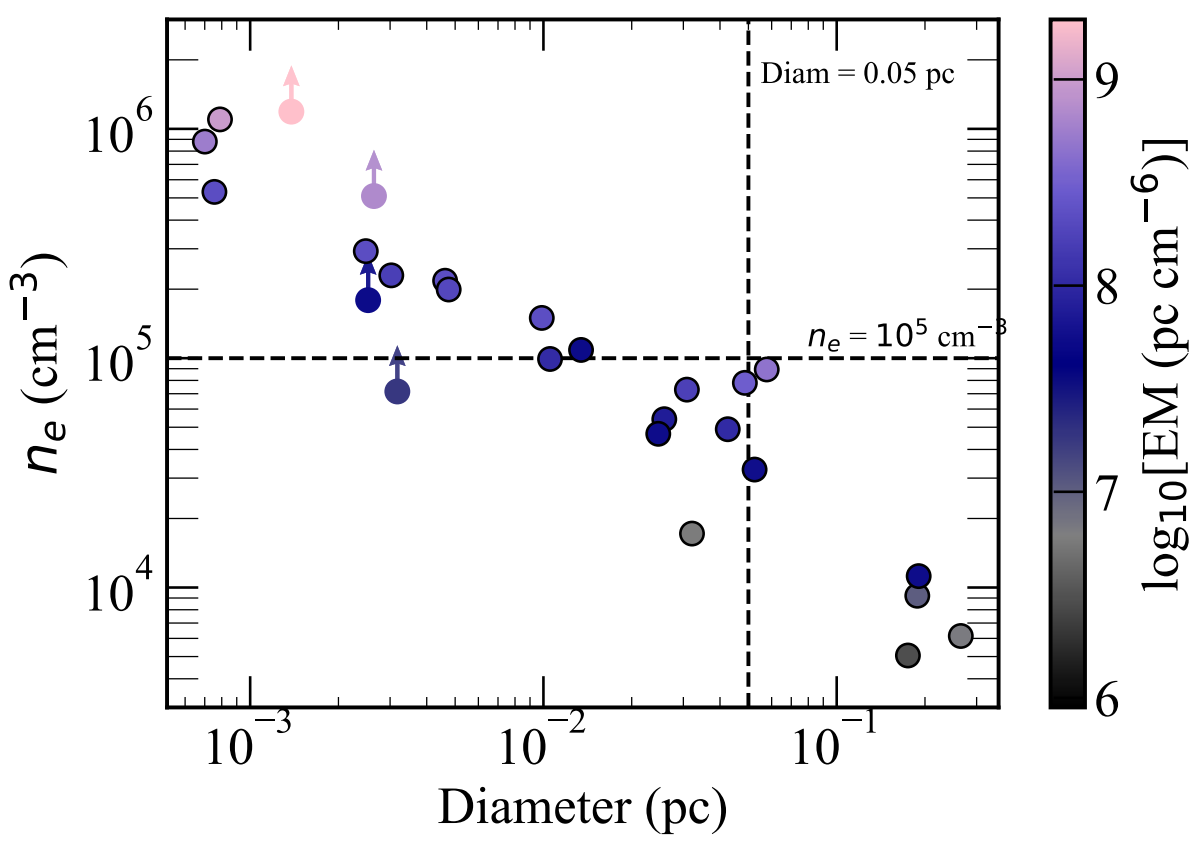

**Figure 10.** Distribution of the physical properties of the 26 radio sources identified in this work. The colour of the data points represents the emission measure of the radio source. Filled circles with upward-pointing arrows denote upper limits for $n_e$ and emission measure, indicating optically thick radio sources. Vertical and horizontal dotted lines indicate the HC Hɪɪ region cutoff.

$$\left(\frac{\nu_t}{\text{GHz}}\right) = 0.3045 \left(\frac{T_e}{10^4\text{K}}\right)^{-0.643} \left(\frac{EM}{\text{cm}^{-6}\,\text{pc}}\right)^{0.476} \tag{4}$$

By considering a typical 20 per cent uncertainty for the diameter and a 20 per cent uncertainty for the emission measure we find that this implies uncertainties of 30 per cent for $\nu_t$ (Yang et al. 2021). The top panel of Figure 9 presents the distribution of the turnover frequency for the 20 Hɪɪ regions, where their physical parameters were obtained through the SED model. The distribution of our sample peaks at ~8 GHz and has a mean of ~9 GHz, which is significantly higher than the expected value of ~5 GHz for UC Hɪɪ regions (Kurtz & Hofner 2005).

### 3.3.2 Lyman continuum Flux

The number of Lyman continuum photons $N_{Ly}$ per second can be calculated using the integrated flux density and kinematic distance for an optically thin Hɪɪ region using (Panagia 1973; Carpenter et al. 1990; Urquhart et al. 2013b):

$$\left(\frac{N_{Ly}}{s^{-1}}\right) = 8.9 \times 10^{40} \left(\frac{S_\nu}{\text{Jy}}\right) \left(\frac{d}{\text{pc}}\right)^2 \left(\frac{\nu}{\text{GHz}}\right)^{0.1} \tag{5}$$

The distribution of the derived $N_{Ly}$ is shown in the lower panel of Fig. 9 and ranges from $10^{44.8}$ to $10^{49.2}$ $s^{-1}$. Assuming that a single zero-age main-sequence (ZAMS) star is the primary source of the ionizing photons, we estimate the spectral type of the radio detections, which are consistent with ZAMS stars between B2 and O5 (Panagia 1973). The estimated uncertainty in the derived $N_{Ly}$ flux is ~20 per cent considering the error in the kinematic distance and flux densities (e.g., Urquhart et al. 2013b; Sánchez-Monge et al. 2013). For optically thick compact Hɪɪ regions, it is likely that the $N_{Ly}$ flux is underestimated, these sources are presented as lower limits (pink region of Fig. 9).

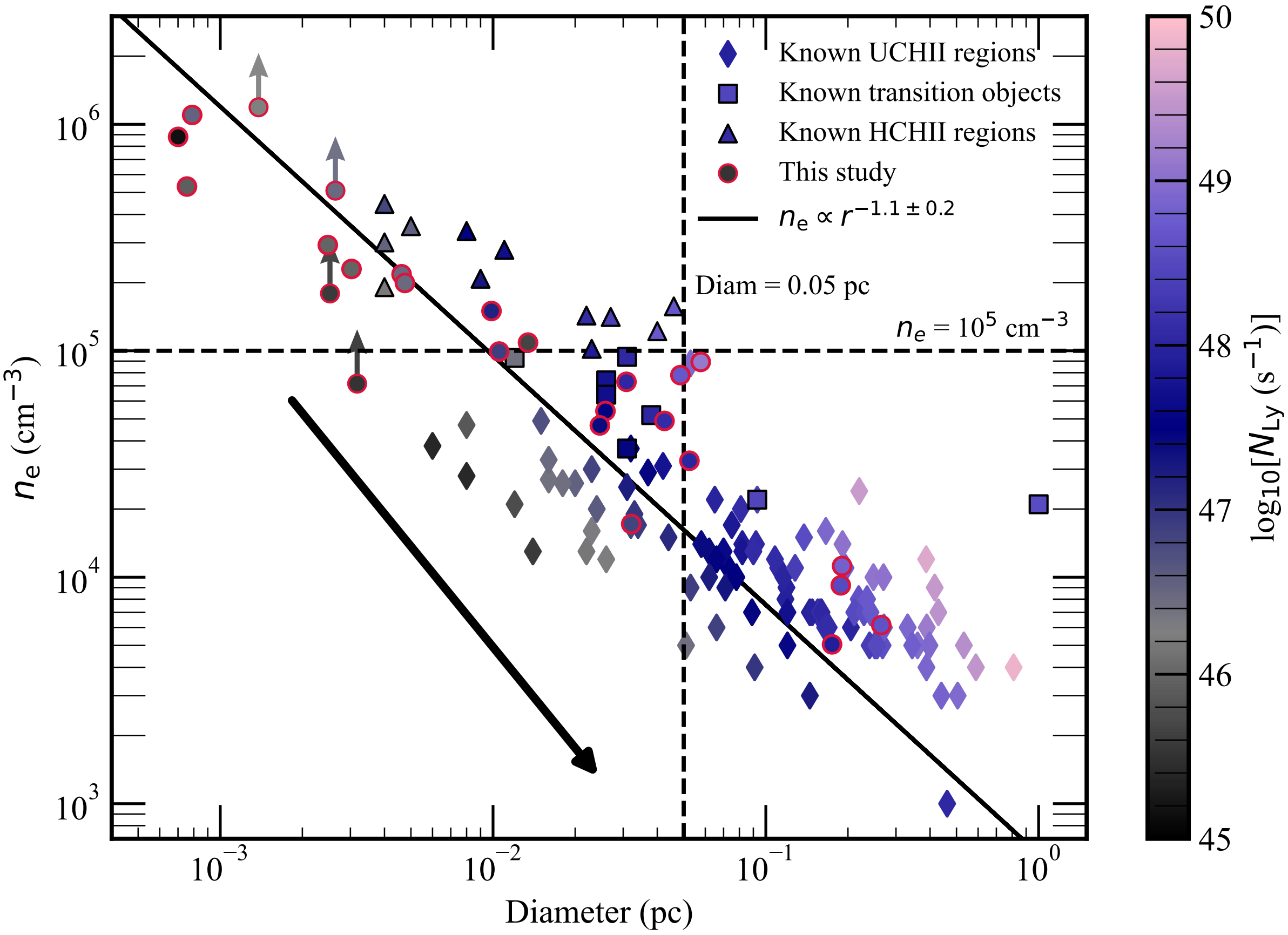

**Figure 11.** Distribution of the physical properties of the 26 radio sources identified in this work, plus HC HII and UC HII regions identified in Kalcheva et al. (2018); Yang et al. (2019); Yang et al. (2021). The colour of the data points denotes Lyman-continuum output ($N_{Ly}$). The diamonds show known UC HII regions while the triangle represent known HC HII regions. The filled circles present the radio sources from this study where we use limits to show the optically thick radio sources. The black solid arrow shows the typical evolutionary trend a HC HII region is likely to follow as these objects expand while the solid black line represents the fitted evolutionary trend of expanding HII regions.

**Table 6.** Quantitative criteria of the physical parameters for HC HII regions, intermediate objects between the two stages and UC HII regions from the literature.

| Parameters | Size (pc) | $n_e$ (cm$^{-3}$) | EM (pc cm$^{-6}$) |
|---|---|---|---|
| HC HII region | < 0.05 | > $10^5$ | > $10^8$ |
| Intermediate objects | 0.05 – 0.1 | $10^4$ – $10^5$ | $10^7$ – $10^8$ |
| UC HII region | > 0.1 | < $10^4$ | < $10^7$ |

### 3.4 Distinguishing between HC HII and UC HII regions

In the previous section, we presented the physical parameters for 26 compact HII regions. In Table 6, we summarise the criteria for HC HII, UC HII regions, and the intermediate objects between these stages (e.g., Wood & Churchwell 1989; Kurtz et al. 1994; Gaume et al. 1995; Kurtz & Hofner 2005; Hoare et al. 2007; Yang et al. 2021).

In Figure 10, we show the distribution of the electron density, diameter and emission measure for the 26 HII regions for which we have accurate distances. The sources that are optically thick in this work are represented as lower limits (arrow pointing upwards) of the electron density while the colours of the data points illustrate the emission measure of each HII region (see the vertical colour bar). On this plot we indicate the region of parameter space we expect to find HC HII regions (i.e., $n_e > 10^5$ cm$^{-3}$ and diam < 0.05 pc).

We find that 14 sources satisfy the physical conditions required for HC HII regions; this includes the four optically thick HII regions. A further 6 sources satisfy the size criterion for HC HII regions but have lower electron densities, these sources are considered to be in a stage between HC HII regions and UC HII regions. The remaining 6 sources, mainly located in the bottom-right quadrant, are considered to be UC HII regions and contribute ∼20 per cent of our sample.

## 4 DISCUSSION

### 4.1 Comparison with known population of compact HII regions

In the preceding section, we have outlined the physical parameters of our sample and identify 14 HC HII regions, 6 objects that are likely to be in a transition stage between HC HII and UC HII regions and 6 UC HII regions. In addition to this, all of the objects in our sample are embedded towards the centres of dense molecular clumps traced by ATLASGAL and are associated with either methanol maser emission or 5-GHz radio emission, all of which are signposts of high-mass star formation. The observed $N_{\rm Ly}$ fluxes support the classification of these objects as ZAMS early B or late O type stars. In this section, we provide a comparison between our sample and the known population of HC Hɪɪ and UC Hɪɪ regions.

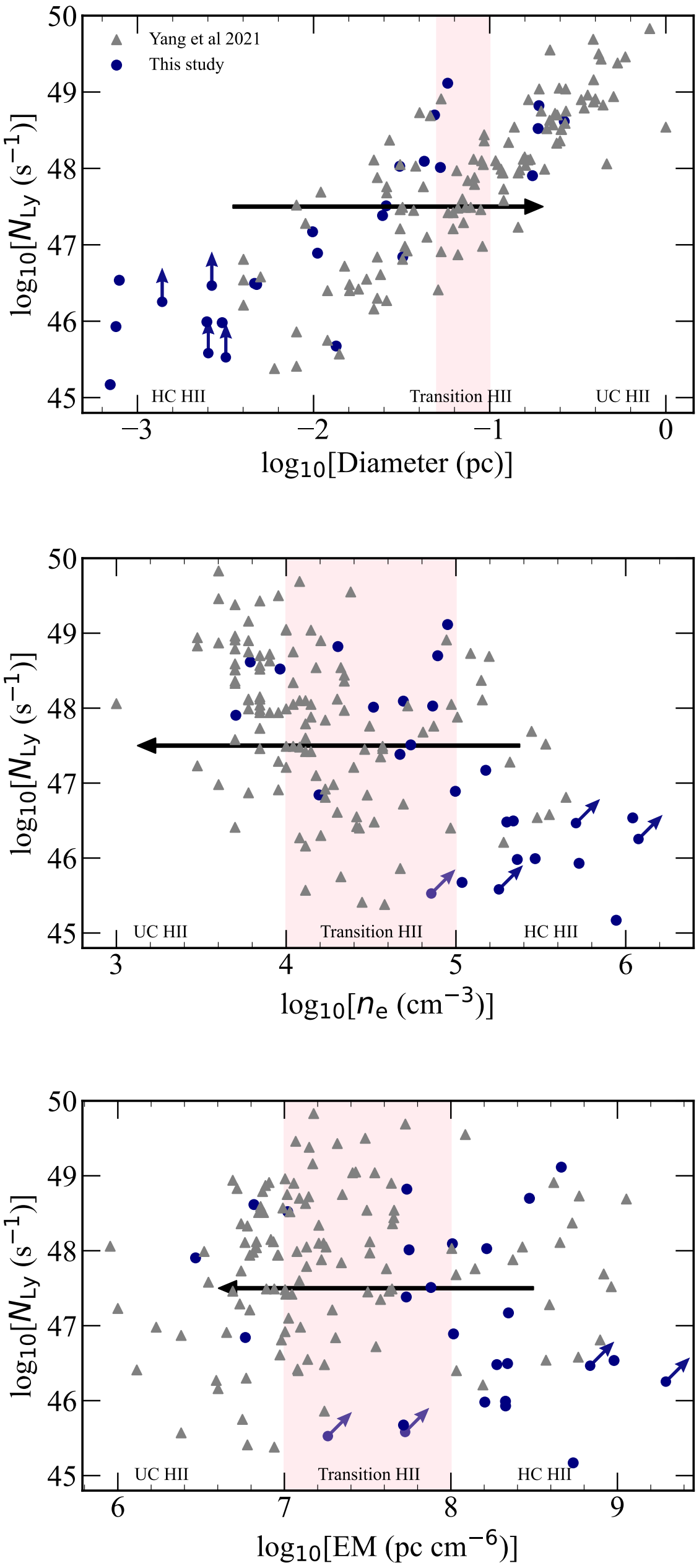

**Figure 12.** Correlation and evolution of the derived physical parameters.The stages of Hɪɪ regions are illustrated at the lower section of the plot, with the transition HII regions depicted by the shaded pink region.

Many theoretical models such as Mezger & Henderson (1967) and Dyson & Williams (1997) have proposed that Hɪɪ regions evolve by expanding continuously over time. Their expansion is due to the ionization front moving into the ambient medium faster than the dynamical time of the gas. When the expansion roughly equals the speed of sound $\sim$10 kms$^{-1}$, thermal-pressure dominates the flow of the ionized gas, precluding accretion (Spitzer 1978; Dyson & Williams 1980). In this discussion we assume that the star has reached the main sequence and is no longer undergoing accretion, and thus the $N_{\rm Ly}$ flux is constant.

In Figure 11, we present the distribution of our sample, along with the known HC Hɪɪ and UC Hɪɪ regions identified in Yang et al. (2021), as a function of their diameter and electron density. The colours of the data points correspond to the $N_{\rm Ly}$ flux, as shown in the colour bar. The HC Hɪɪ, UC Hɪɪ, and transition objects identified in this work, and the known HC Hɪɪ and UC Hɪɪ regions from the literature, are in good agreement with each other, and with the evolutionary models discussed above. The solid black arrow represents the characteristic behaviour of the Strömgren sphere expansion which assumes uniform density and temperature. This illustrates the expected evolutionary trend from HC Hɪɪ to compact Hɪɪ regions with the relationship $n_{\rm e} \propto r^{-1.5}$ where $r$ = diameter/2. Observations have shown that Hɪɪ regions do not expand spherically due to density inhomogeneities in the surrounding gas, which are likely to impede their expansion. Our sample and the known population of compact Hɪɪ regions follows the relationship $n_{\rm e} \propto r^{-1.1\pm0.2}$, which, within the uncertainties, provide a better constraint for the expansion of Hɪɪ regions. This relationship is illustrated by the solid line in Figure 11.

To further investigate how the individual physical parameters change as a Hɪɪ region evolves, we present the Lyman continuum output as a function of $n_{\rm e}$, diameter and emission measure for our sample and the known populations of young Hɪɪ regions in Fig. 12. The optically thick sources are presented as upper limits to the $N_{\rm Ly}$, $n_{\rm e}$, and emission measure. The solid black arrow represents the expected evolutionary trend of the physical characteristics.

In the upper and middle panel of Fig. 12, we present $N_{\rm Ly}$ as a function of diameter and $n_{\rm e}$, respectively. The relationship between the diameter and the $n_{\rm e}$ of the Hɪɪ region is inversely proportional, indicating that the smallest and densest Hɪɪ regions are likely to surround early B type stars. It is interesting to note that the compact HC Hɪɪ regions (diam < $10^{-2}$ pc) in this work only surround early B-type stars (i.e., $N_{\rm Ly}$ < $10^{47}$ s$^{-1}$), whereas the larger HC Hɪɪ regions (diam > $10^{-2}$ pc) surround late O-type stars (i.e., with $N_{\rm Ly}$ > $10^{48}$ s$^{-1}$). This could be attributed to the rapid expansion of Hɪɪ regions around late O-type stars, driven by their high luminosity and strong ionizing photon flux that results in these types of high-mass stars being associated with lower gas density at larger radii. Additionally, there is a region of parameter space (See the lower right corner of the upper panel and the lower left corner of the middle panel in Fig. 12) where we do not find any UC Hɪɪ regions that surround early B-type stars. This is likely to be due to either sensitivity as the surface brightness decreases as they expand or due to the nature of the interferometric observations that filter out sources above a certain angular scale.

In the lower panel of Fig. 12, we present the distribution of the $N_{\rm Ly}$ flux with respect to the emission measure. We find no correlation between the $N_{\rm Ly}$ flux and the emission measure of evolving Hɪɪ regions, suggesting that the emission measure is invariant with respect to $N_{\rm Ly}$. A similar trend has been reported in the works of Yang et al. (2021), who confirm that there is effectively no relationship between $N_{\rm Ly}$ and emission measure (e.g., see Sect 5.1 from Yang et al. 2021).

The HC Hɪɪ regions detected in this study exhibit a distinctive characteristic whereby they tend to be associated with lower bolometric-luminosity stars. In Figure 13, we show cumulative distribution functions for the bolometric luminosity and $N_{\rm Ly}$ flux for the HC Hɪɪ regions identified in the literature and the HC Hɪɪ regions confirmed in this work. Upon initial inspection, the distributions ap-pear to be different. The grey dotted line in Fig. 13, represents the median of our sample, while the navy dotted line indicates the median value of Yang et al. (2021) sample. It appears that the majority of our sample tend to surround early B-type stars whereas the HC HII from Yang et al. (2021) sample surround late O-type stars. The difference in median values confirms there is a substantial distinction between the two HC HII region samples. To determine if there are similarities in the bolometric luminosities and Lyman fluxes between the two samples of HC HII regions, we use a two-sample Kolmogorov–Smirnov (KS) test. Our null hypothesis states that the two samples of HC HII regions are drawn from the same underlying distribution; this can be rejected if the $p$-value is lower than $3\sigma$ (i.e., < 0.0013). We find that the distribution of the Lyman continuum flux is significantly different, while the bolometric luminosity shows a marginal difference of $\sim 2.8\,\sigma$.

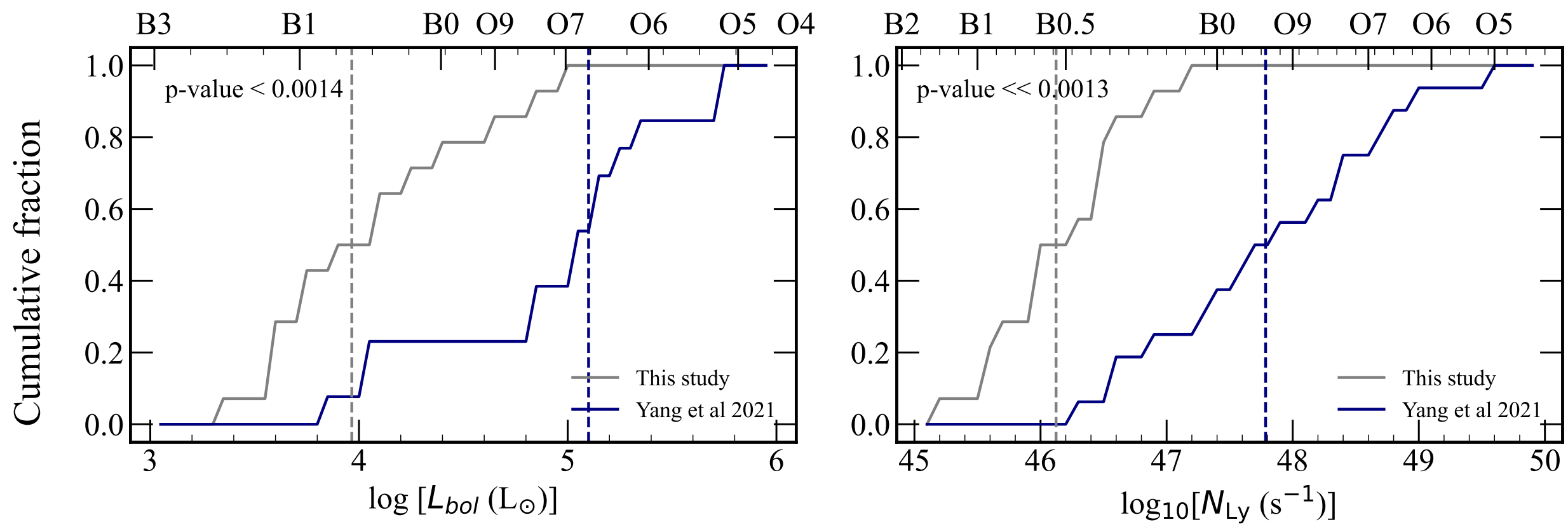

**Figure 13.** Cumulative distribution functions for bolometric luminosity (left panel) and $N_{Ly}$ flux (right panel) for the sample of HC HII regions identified in this work and in the literature (Yang et al. 2021). The dotted line represents the median value of each sample.

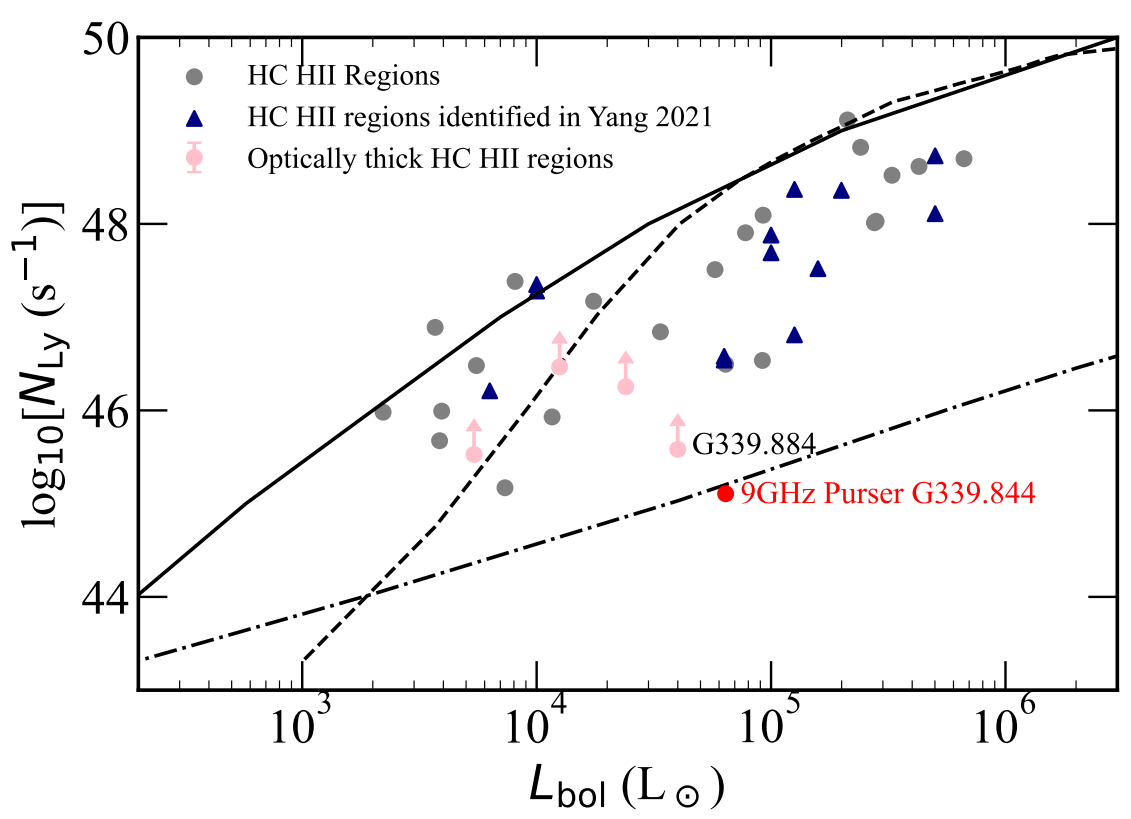

**Figure 14.** Lyman-continuum photon flux as a function of bolometric luminosity for the HC HII regions identified in this work and in the literature. The grey and pink circles show our sample of HC HII and optically thick HC HII regions, respectively. The blue triangles represent the HC HII and transition objects identified in Yang et al. (2021). The dashed line indicates the relationship for OB ZAMS-star models (Martins et al. 2005; Davies et al. 2011) and the dot-dashed line represents a fit to radio jets found in Purser et al. (2016, 2021). The solid black line corresponds to the Lyman continuum expected from a blackbody with the same radius and temperature as a ZAMS star (see Sánchez-Monge et al. 2013 for more detail).

In Fig 14 we show the Lyman photon flux as a function of bolometric luminosity for both samples of HC HII regions. The distribution further confirms that the HC HII regions identified in this work surround lower-luminosity stars. A possible explanation for this could be that the HC HII regions reported in previous literature were identified from a lower-frequency survey that is sensitive primarily to the brightest HC HII regions, due to the fact that they are optically thick at these frequencies. Whereas our sample of HC HII regions is targeted towards sites of methanol masers that are not intrinsically bright at higher resolutions. In addition, high-frequency radio observations are more sensitive to lower-luminosity high-mass stars.

Nonetheless, as demonstrated by Fig. 14, the HII regions within our sample align well with the broader population of HII regions. All of the objects are located significantly above the jet line (see dot-dashed line on Fig. 14), apart from G339.884, which will be discussed later. This alignment provides strong evidence of their association with massive stars and rules out the possibility of them being radio jets, which tend to have lower radio luminosities compared to their bolometric luminosity.

### 4.2 Physical properties of the optically thick HII regions

We have identified four HC HII regions that remain optically thick ($\alpha > 0.5$) and compact (diam < 0.05 pc) at 24 GHz. These are likely to be the youngest HII regions in our sample. The radio sources are all within 1 arcsec of the methanol-maser position. Given the 1-arcsec positional uncertainty in the maser positions (Green et al. 2010), we consider them coincident for this discussion. Considering that methanol masers often reside in the dusty molecular envelope (Urquhart et al. 2015), they are sensitive probes for discovering the earliest stages of massive star formation (Breen et al. 2011). Methanol masers are pumped by infrared radiation and require high temperatures and densities to be stimulated (Sobolev & Deguchi 1994). These conditions are likely to be found between 100–1000 au from the high-mass protostellar object and are signposts of ongoing disk accretion (Breen et al. 2013).

Other molecular masers (i.e., $H_2O$ and OH) can provide valuable insights into the environment, kinematics and feedback processes associated with massive star formation. Water masers are collisionally pumped and the conditions required to excite them are

**Table 7.** Physical characteristics of the optically thick sources and their embedding environments. The typical error in the ATLASGAL clump properties are estimated to be of order 20 per cent.

| Radio name | CSC name | ATLASGAL Clump Properties | | | | Radio Derived Properties | | | | |
|---|---|---|---|---|---|---|---|---|---|---|
| | | Distance (kpc) | Mass ($10^2$ M$_\odot$) | Luminosity ($10^5$ $L_{bol}$) | Temperature (K) | Diameter ($10^{-3}$ pc) | $n_e$ ($10^4$ cm$^{-3}$) | $EM$ ($10^7$ pc cm$^{-6}$) | $\log_{10}$ [$N_{Ly}$] (s$^{-1}$) | ZAMS Spectral type |
| G326.475+0.703 | AGAL326.474+00.702 | 2.6 | 8.4 | 5.4 | 20.7 | 3.2 | 7.2 | 1.8 | 45.5 | B1 |
| G332.987−0.487 | AGAL332.986−00.489 | 3.5 | 40[a] | 24[c] | 24.6[a] | 1.4 | 118.9 | 195.3 | 46.3 | B0.5 |
| G333.029−0.063 | AGAL333.029−00.061 | 2.8 | 2.5 | 12.5 | 30.8 | 2.6 | 50.9 | 68.6 | 46.5 | B0.5 |
| G339.884−1.259 | AGAL339.884−01.257 | 2.1 | 4[b] | 40[d] | – | 2.5 | 17.9 | 5.3 | 45.5 | B1 |

Note: references. [a] He et al. (2016), [b] Liu et al. (2019), [c] Wheelwright et al. (2012), [d] Zhang et al. (2019).

high densities (~$10^7$ cm$^{-3}$) and temperatures up to 500 K (Elitzur et al. 1992). These conditions are exclusively found in the shocked gas associated with molecular outflows (Claussen 2002). Hydroxyl (OH) masers are also collisionally pumped and are widely recognised to trace shocked gas in the interface between neutral hydrogen and molecular hydrogen known as photo-dissociation regions (PDR) (Frail & Mitchell 1998). OH masers are thought to trace a slightly more evolved evolutionary stage than the class II $CH_3OH$ masers (Beuther et al. 2019).

In the following section, we combine our findings with results from the literature to build a more comprehensive picture of the environments in which these sources are embedded. Detailed characteristics of the clump are provided in Table 7 and images of their infrared environment are shown in Figs. 15–18. Given that these are optically thick, the $N_{Ly}$ photon flux and the derived ZAMS spectral type for these sources are lower limits.

**G326.475+0.703 (Figure 15):** The high-frequency radio emission is embedded towards the centre of the dense molecular clump AGAL326.474+00.702 (Urquhart et al. 2014b), which is itself associated with the RCW 95 massive star-forming region (Rodgers et al. 1960). This particular clump has been classified as an infrared dark cloud (i.e., SDC G326.476+0.706; Peretto & Fuller 2009) and has a mass of 840 M$_\odot$ and volume density of $10^6$ cm$^{-3}$ (Urquhart et al. 2014b). Inspection of the IRAC mid-infrared images reveals the presence of a bright source to the south-east but this is located on the periphery of the central part of the clump, which is mid-infrared quiet at 8 and 24 $\mu$m.

High-resolution (~ 4-arcsec) observations of the dust emission at 350 GHz ($\lambda \sim 0.9$ mm) by Csengeri et al. (2017a) reveals the presence of three dense cores (~$10^7$ cm$^{-3}$) located towards the centre of the clump. These are classified as 326.4745+0.7027-MM1, MM2, and MM3, with masses of 150 M$_\odot$, 27 M$_\odot$, and 20 M$_\odot$, respectively. The HC Hii region and the methanol maser are both positionally coincident with the centre of the most massive core MM1, which is itself located at the centre of the clump. At a distance of 2.6 kpc, the angular separation between the HC Hii region, methanol maser and the centre of the core put them within approximately 1550 au of each other, roughly corresponding to the size of a typical high-mass protostellar accretion disk taking into account the uncertainties in the positions.

Analysis by Cyganowski et al. (2008) has revealed the presence of two extended green objects (EGOs) that are located to the east and west of the central core. The positions of these are shown in Fig. 15 as red crosses. EGOs are identified by an excess in the 4.5-$\mu$m IRAC band and is thought to be the result of strong emission from the rotationally excited $H_2$ ($\nu$ = 0 − 0) and CO ($\nu$ = 1 − 0) band-head emission (c.f. Fig. 1 of Reach et al. 2006) that fall within this waveband. The excited $H_2$ results from shocked gas associated with outflows while the CO bandhead emission at 2.3 $\mu$m is thought to trace warm (T = 2500−5000 K) and dense (~$10^{11}$ cm$^{-3}$) gas Chandler et al. (1993) and is thought to be associated with accretion disks. Given that these two EGOs appear approximately 8 to 10 arcsec away from the central core and outside of the lowest contour of the ALMA map, suggests that the excess emission could be due to the presence of a bipolar molecular outflow rather than accretion. However, analysis of the of the ALMA CO (3-2) data from SPARKS does not reveal an outflow in the direction of the northern EGO (EGO G326.47+0.70) and so the true nature of the emission is less certain. A detailed analysis of these data will be described in a future publication.

Although there is no significant 8- or 24-$\mu$m emission, there is strong 70-$\mu$m emission seen in the HiGAL image (Molinari et al. 2010), resulting in this source being classified as protostellar in Urquhart et al. (2022). The association of the EGOs and presence of a methanol maser strongly suggests that a high-mass protostar is forming within the core and that active accretion is ongoing. Furthermore, the presence of the radio emission would further indicate that the high-mass protostellar object has recently arrived on the main sequence and began ionizing its natal environment. The Lyman flux and bolometric luminosity are consistent with the presence of a high-mass ZAMS star with a spectral type of B1 or earlier, corresponding to a stellar mass of more than 10 M$_\odot$.

**G332.987−0.487 (Figure 16):** This object is compact, with an angular size similar to that of the beam (~0.4 arcsec) and is embedded towards the centre of the dense molecular clump (AGAL332.986−00.489; Contreras et al. 2013). This clump is embedded within the G333 giant molecular cloud, which is located at a distance of 3.6 kpc (Lockman 1979) and is itself the fourth most active star-forming region in the Galaxy (Urquhart et al. 2014a). Observations by Billington et al. (2019) using $NH_3$ (1,1) and (2,2) have identified a high-density region towards the centre of the ATLASGAL clump that has an approximate size of 0.2 pc. The radio emission is coincident with the centre of both $NH_3$ and ATLASGAL clumps and the methanol maser G332.987−00.487 (Caswell 2010) which is offset by 0.7 arcsec east of the radio emission.

The mid-infrared environment reveals the presence of a bright 8-$\mu$m point source associated with an EGO (i.e., extended 4.5-$\mu$m emission), which is slightly offset from the peak of the dust emission. This has been classified as a MYSO in the Red MSX Source (RMS) survey by Lumsden et al. (2013) and more recently by ATLASGAL in Urquhart et al. (2022). High-resolution near-infrared spectroscopy of the CO ($\nu$ = 2 − 0) bandhead emission towards a sample of MYSOs that includes this source were conducted by Ilee et al. (2013) using the Very Large Telescope (VLT). CO bandhead emission at 2.3$\mu$m traces warm (T = 2500−5000 K) and dense (~$10^{11}$ cm$^{-3}$) gas. These conditions are consistent with the environment typically found in the inner regions of accretion discs (Wheelwright et al. 2012; Ilee et al. 2013). The presence of the CO bandhead and MMB methanol maser indicates ongoing accretion through a disk surrounding the MYSO. Ilee et al. (2013) assumed a geometrically thin disk and used a model to determine the disk properties surrounding the associated MYSO. An analytical model of the temperature and density was used to fit the individual line profiles between 2.296 and 2.298 $\mu$m. The mass and effective temperature determined by Ilee et al. (2013) for MYSO G332.9868−00.4871 is 16.6 $M_{\odot}$ and 30,000 K, respectively. These values are consistent with the MYSO having a ZAMS spectral type of B0.5, which is also consistent with the ionizing photon flux derived from the radio data.

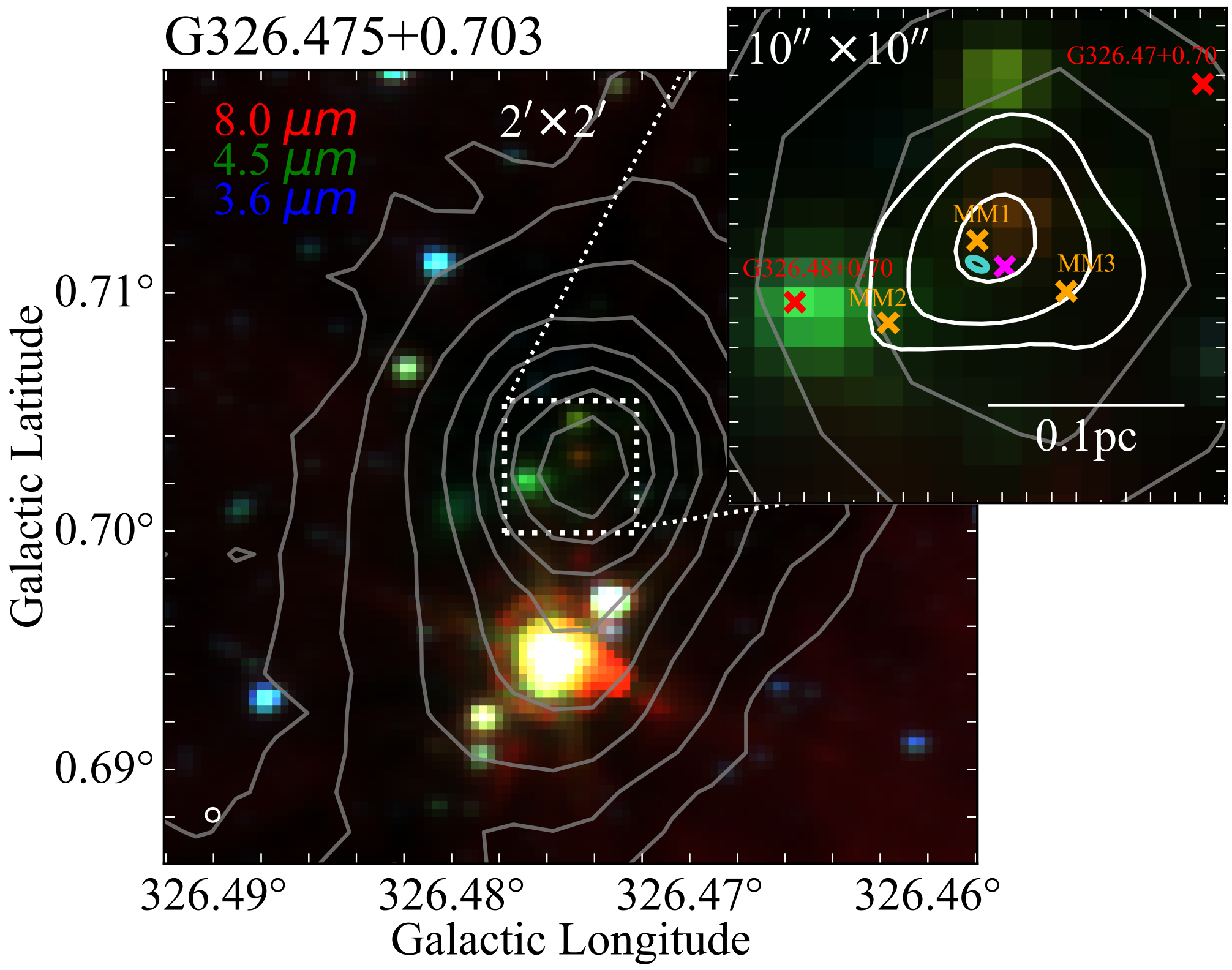

**Figure 15.** Three-colour composite images from Spitzer GLIMPSE 8-$\mu$m (red), 4.5-$\mu$m (green) and 3.6-$\mu$m (blue) bands for HC HII region G326.475+0.703. In the upper-right zoomed panel, the cyan ellipse shows the high-frequency (24-GHz) radio source and the magenta crosses show the position of the MMB methanol maser. The 870-$\mu$m dust emission from ATLASGAL is traced by the grey contours. The white contours show the 350-GHz dust continuum emission from SPARKS where the orange crosses present the positions of the dense cores as identified in Csengeri et al. (2017a). The red cross show the peak position of EGOs as identified in Cyganowski et al. (2008).

The angular offset between the HC HII region and MYSO G332.9868−00.4871, that is associated with the accretion disk, is ~0.3 arcsec, given the positional uncertainties these are considered to be associated with the same object.

**G333.029−0.063 (Figure 17):** This very compact source is located towards the centre of dense molecular clump AGAL333.029−00.061 (Contreras et al. 2013), which is also embedded within the G333 giant molecular cloud (Lockman 1979). The mid-infrared environment reveals multiple point sources, indicating the presence of a small protocluster located within the central part of the clump, but offset from the peak of the dust emission, which is mid-infrared dark. The high-frequency radio emission is coincident with the edge of the brightest of the mid-infrared sources, which is extended and located to the south of the centre of the clump. High-resolution (~ 2.5-arcsec) continuum maps at 5 GHz produced by CORNISH-south (Irabor et al. 2023) reveal the presence of an unresolved point source, G333.0290−0.0631 that is offset by 0.5 arcsec from the high-resolution source, however, the resolution of the CORNISH observations is a few arcsec and given the uncertainties, the radio emission is likely to be associated with the same star. The extended nature of the 8-$\mu$m emission and its association with the 5-GHz radio continuum emission has led to this source being classified as an HII region in the RMS survey (Lumsden et al. 2013).

Methanol, $H_2O$ and OH maser species have been detected within 3 arcsec of the HC HII region. The $H_2O$ maser is located north-east of the methanol maser, which is coincident with the 5-GHz radio continuum emission (offset ~0.3 arcsec). Two 1720-MHz OH masers have been recently identified towards this region by Qiao et al. (2020) as part of the Southern Parkes Large-Area Survey in Hydroxyl (SPLASH) project (Dawson et al. 2014). These are identified as G333.029−0.063-1665A and G333.029−0.063-1720A and are shown as blue crosses in Fig. 17. The tight correlation in position between maser species strongly suggests that the high-mass protostellar object is still undergoing accretion. The presence of the $H_2O$ maser confirms that there is shocked gas, which is likely to originate from outflows. Furthermore, considering 1720-MHz OH masers trace shocked gas towards the edge of an expanding HII region bubble, it is likely that the embedded high-mass protostar has recently begun ionizing its surroundings.

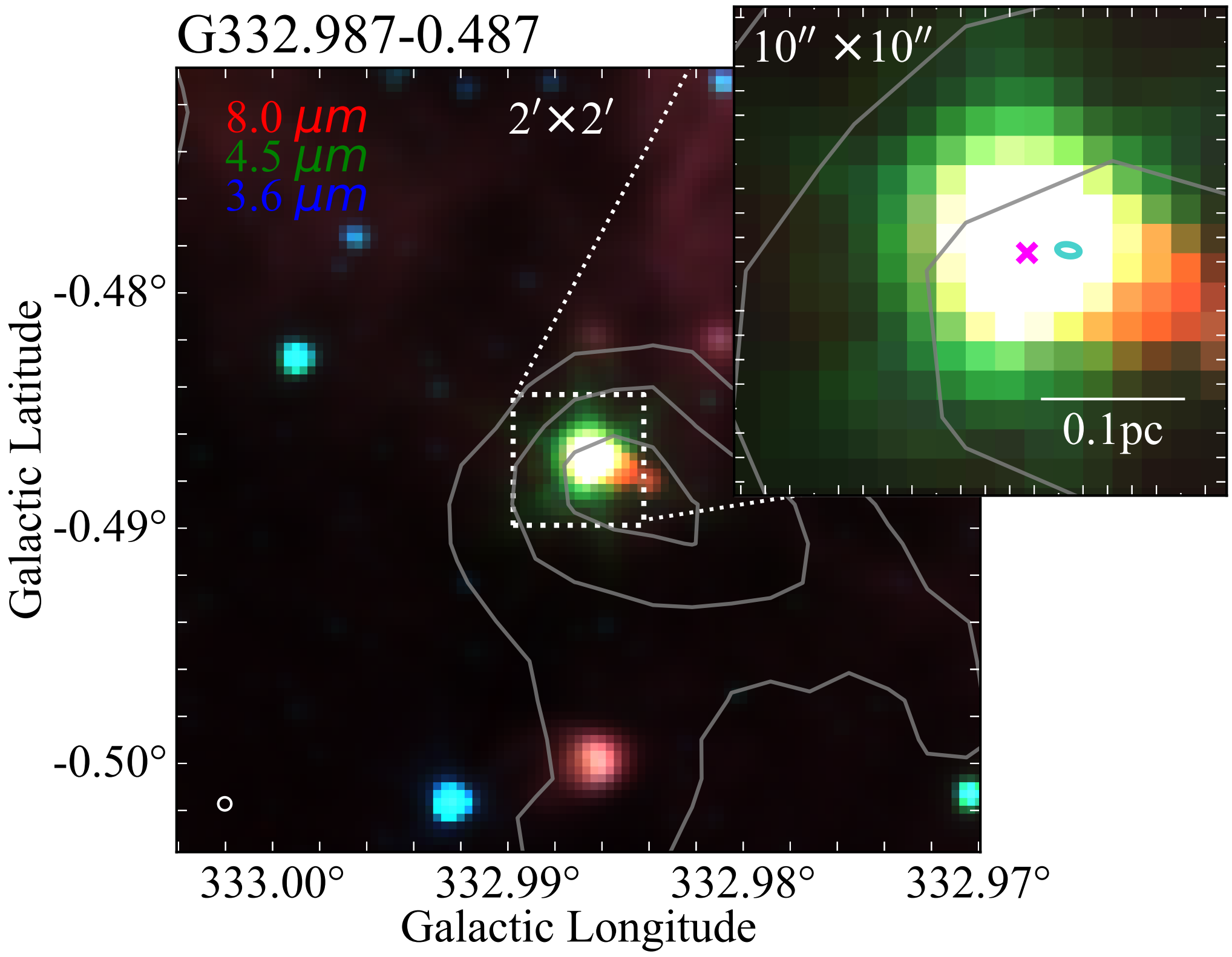

**Figure 16.** Three-colour composite images from Spitzer GLIMPSE 8-$\mu$m (red), 4.5-$\mu$m (green) and 3.6-$\mu$m (blue) bands for HC H II region G332.987−0.487. The 870-$\mu$m dust emission from ATLASGAL is traced by the grey contours. In the upper-right zoomed panel, the cyan ellipse shows the high-frequency (24-GHz) radio source and the magenta cross shows the position of the MMB methanol maser.

**G339.884−1.259 (Figure 18):** The high-frequency radio emission is detected toward the centre of dense molecular clump AGAL339.884−01.257 (Contreras et al. 2013) and is located at a distance of 2.1 ± 0.4 kpc (Krishnan et al. 2015). High-resolution (~0.06-arcsec) images made with the mid-infrared spectrometer OSCIR at the Keck Observatory have resolved the mid-infrared emission into three distinct peaks (De Buizer et al. 2002); these are identified as 1A, 1B, and 1C, all of which lie within a region of ∼3000 au. High-resolution (∼0.3-arcsec) ALMA submillimetre observations of dust emission at 220 GHz (1.3 mm) by Zhang et al. (2019) reveal the presence of a dense core that has a gas mass between 58 and 23 $M_\odot$ assuming a temperature range of 30−70 K, respectively. The HC H II region and the methanol maser are both positionally coincident with the centre of the core, which is itself located towards the centre of the ATLASGAL clump. It is interesting to note that the submm peak does not coincide with any of the mid-infrared peaks. Instead, it appears to be situated in the gap between mid-infrared points B and C, and is therefore mid-infrared quiet at 10 and 18 $\mu$m (De Buizer et al. 2002).

Molecular masers such as $H_2O$ and OH have been detected within the centre of the core. Two 1720-MHz OH masers are positioned to the south west of the radio emission and are offset by ∼1 arcsec (Caswell et al. 1995). A $H_2O$ maser is located 0.7 arcsec east of the radio source. As previously mentioned, these masers are known to trace the shocked gas within an outflow cavity. The ALMA $^{12}$CO(2−1) observations by Zhang et al. (2019) reveal a large-scale molecular outflow with a northeast–southwest orientation emanating from the centre of the submm emission. These are shown as solid arrows in Fig 18. A secondary, weaker outflow (dotted arrows on Fig 18) has been identified in a southeast–northwest orientation, which roughly coincides with an ionised radio jet identified by Purser et al. (2016) (indicated by green arrows on Fig 18). The orientation of these outflows would strongly suggest the presence of an embedded proto-binary system with individual outflows roughly perpendicular to each other (Zhang et al. 2019).

Comparing our high-frequency flux densities with those reported in the literature, we find that the fluxes recorded in this study are approximately double the values reported in Purser et al. (2016). We verified our calibration and the fluxes derived for the primary flux calibrator and find it to be consistent with expected values within 10 per cent. Since the observations by Purser et al. (2016) were taken in Feb 2011, the differing flux densities could be attributed to the variability of the radio sources, considering the time between the observation sets. The variability of flux densities has been reported in several compact and young H II regions detected by the GLOSTAR survey (e.g., Yang et al. 2023). At a distance of 2.7 kpc and a bolometric luminosity of $6.4 \times 10^4$ $L_\odot$, Purser et al. (2016) uses the 9-GHz radio flux to derive the radio luminosity. They find that the object aligns well with the expected luminosity of a radio jet. However, considering the object is optically thick at 9 GHz it is likely that the radio luminosity is underestimated. As shown in Fig. 14, our 24 GHz radio flux puts this source above the expected radio jet line. Our observations give us a better constraint

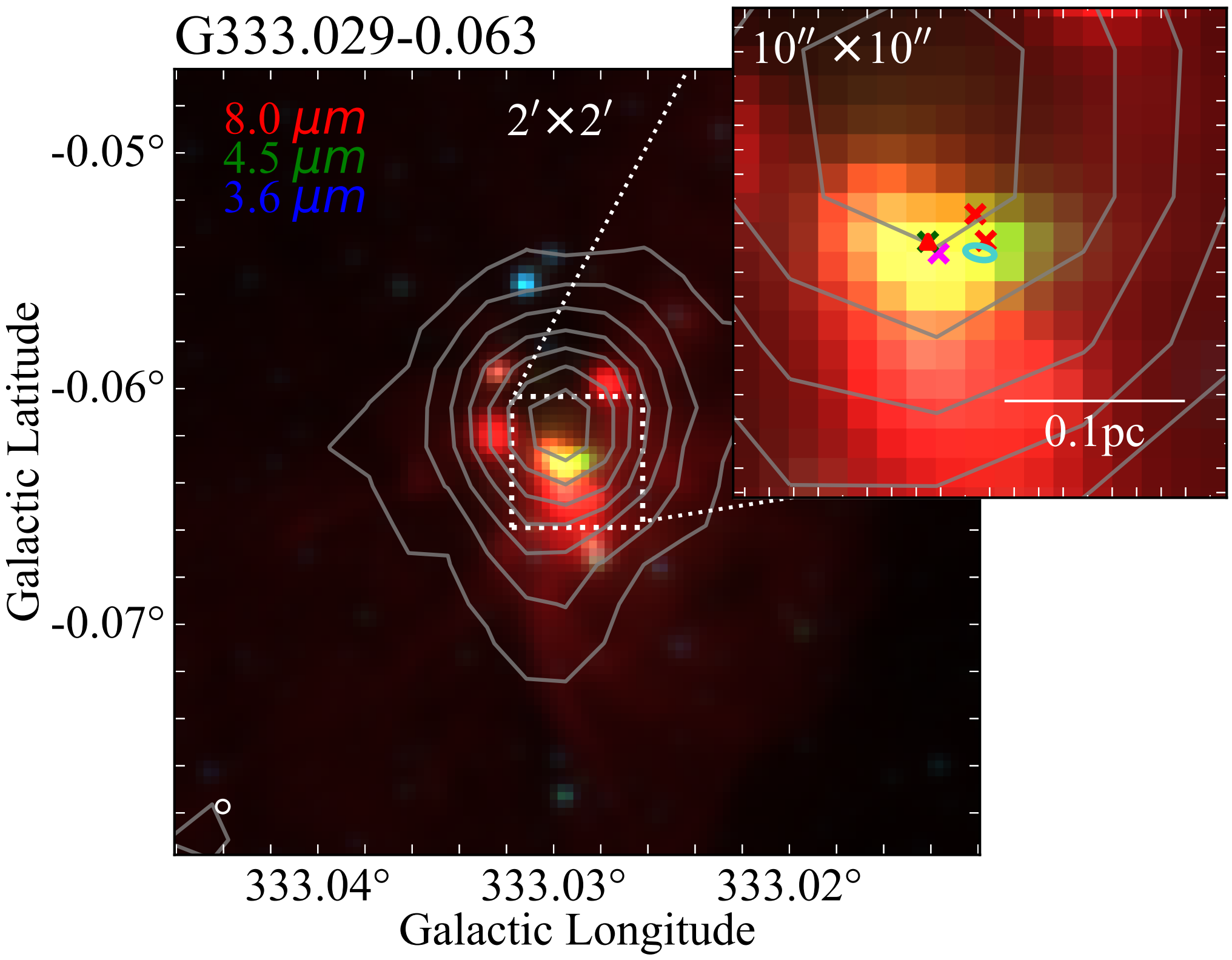

**Figure 17.** Three-colour composite images from Spitzer GLIMPSE 8-$\mu$m (red), 4.5-$\mu$m (green) and 3.6-$\mu$m (blue) bands for HC Hii region G333.029−0.063. The 870-$\mu$m dust emission from ATLASGAL is traced by the grey contours (Schuller et al. 2009). In the upper-right zoomed panel, the cyan ellipse shows the high-frequency (24-GHz) radio source and the magenta cross shows the position of the MMB methanol maser. The 5-GHz CORNISH-South radio source is shown as a red triangle (Irabor et al. 2023). OH and $H_2O$ masers are shown as red and green crosses, respectively. (Qiao et al. 2020; Breen & Ellingsen 2011).

on the lower limit; however, the true nature of this object is still uncertain.

### 4.3 Matches with ALMA cores

The Search for high-mass protostars with ALMA up to 5 kpc (SPARKS) project observed 46 mid-infrared quiet ATLASGAL clumps with ALMA in an effort to characterise objects in the earliest stages of high-mass star formation (Csengeri et al. 2017a). Given that methanol masers are considered to be associated with the earliest stages in high-mass star formation, many of the cores identified by SPARKS host methanol masers: 42 out of the 112 cores (37.5 per cent; Chibueze et al. 2017). Comparing the SPARKS cores with our sample of high-frequency compact radio sources, we find four matches located in three SPARKS fields. One of these (G326.475+0.703) has already been discussed in the previous subsection, the other two radio sources are G338.925+0.556 and G339.681−1.207, which are matched with SPARKS fields G338.9249+0.5539 and G339.6802−1.2090, respectively. Below we provide a summary of the other two matches between SPARKS cores and radio sources.

**G338.925+0.556 (Figure 19):** There are three SPARKS cores associated with the ATLASGAL clump AGAL338.926+00.554. The most massive core (∼250 $M_\odot$, MM1) is located towards the centre of the clump and is mid-infrared quiet. The second most massive core (∼200 $M_\odot$, MM2) is located to the north-east of MM1, and it is this core, that is coincident with a methanol (G338.925+00.557; Caswell 2009), OH maser (OH 338.925+00.557; Sevenster et al. 1997) and compact high-frequency (24 GHz) radio source identified in this work (G338.925+0.556). It is also associated with a compact infrared source that appears to have an excess 4.6 $\mu$m emission, consistent with the presence of shocked gas. The angular offset for the radio, excess 4.6 $\mu$m emission and methanol maser and MM2 are all less than 1 arcsec, which at the distance of 4.2 kpc (Urquhart et al. 2022) corresponds to a physical offset of less than 0.02 pc and so can be considered to be associated with each other. The third SPARKS core (∼60 $M_\odot$, MM3) is coincident with an extended green object (EGO G338.92+0.55(a); Cyganowski et al. 2008) indicating star formation in this core is already underway.

The high frequency radio source is not detected by CORNISH-South at 5 GHz (Irabor et al. 2023) and is also a non-detection at 8.6 GHz (Urquhart et al. 2007a). Using the 3$\sigma$ of these radio maps as upper limits we have fitted the radio SED and this shows the Hii region is likely to be optically thick at 5 GHz with a turnover close to 18 GHz, consistent with this being classified as a HC Hii region. The Lyman flux is $3.4 \times 10^{46}$ photons $s^{-1}$ consistent with being ionised by ZAMS star with a spectral type of B0.5.

To the north of the SPARKS cores (MM3), but still located within the boundary of the ATLASGAL clump is a low surface brightness compact Hii region (G338.9237+00.5618) and there is a high-mass YSO (G338.9196+00.5495) located to the south west of the SPARKS cores; both the Hii region and YSO have been identified by the identified by the RMS survey (Lumsden et al. 2013). The high-mass YSO is associated with an OH maser (OH 338.92+00.55; Caswell et al. 1995) and EGO (EGO G338.92+0.55(b); Cyganowski et al. 2008). The mid-infrared composite image also reveals the presence of several other compact infrared sources coincident with the clump.

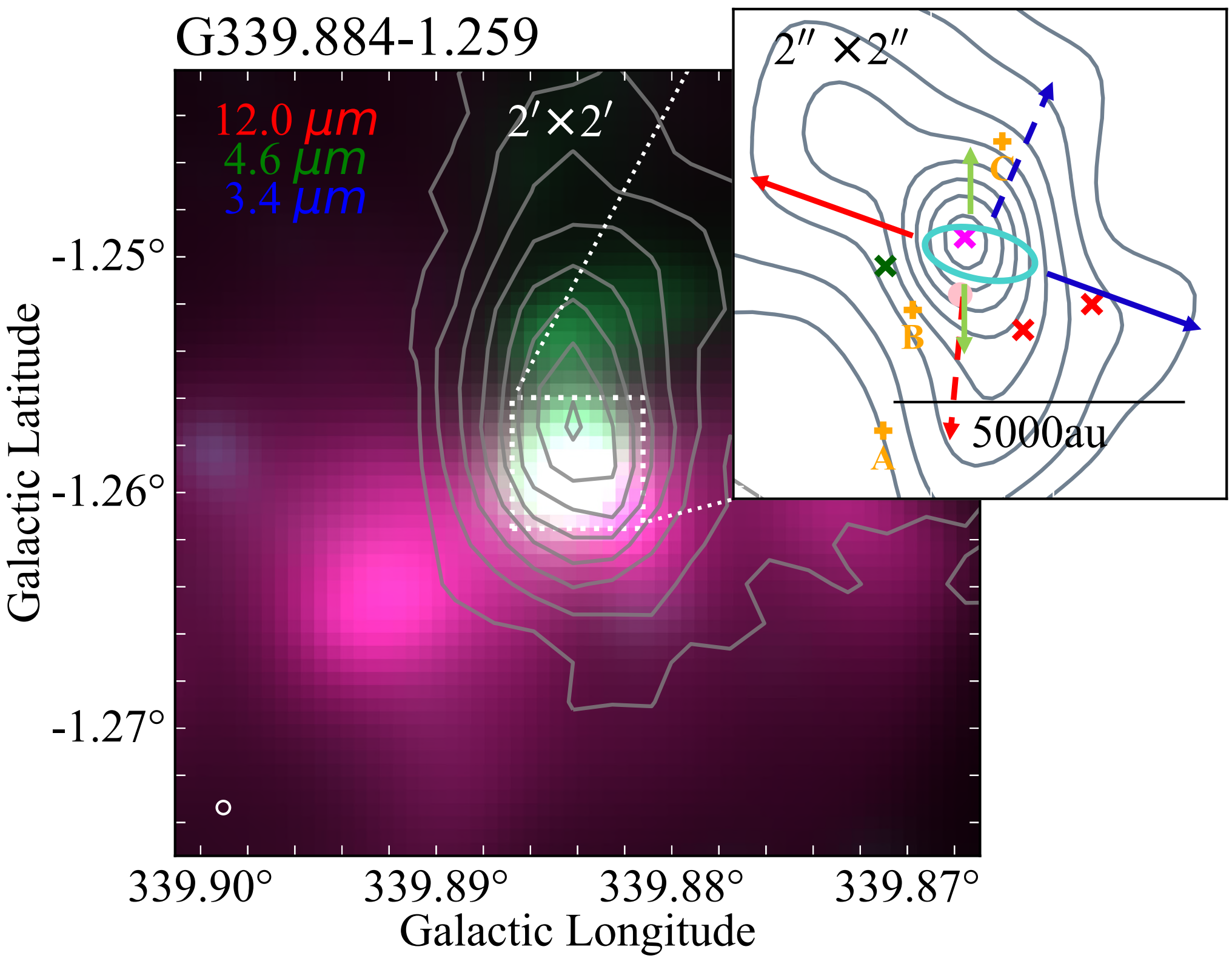

**Figure 18.** Three-colour composite images from WISE 12-$\mu$m (red), 4.6-$\mu$m (green) and 3.4-$\mu$m (blue) bands for HC HII region G339.884−1.25. The 870-$\mu$m dust emission from ATLASGAL is traced by the grey contours. In the upper right zoomed panel, the cyan ellipse shows the position of the high-frequency (24-GHz) radio source and the magenta cross shows the position of the MMB methanol maser. OH and $H_2O$ masers are shown as red and dark green crosses, respectively (Qiao et al. 2020; Breen & Ellingsen 2011). The 10-$\mu$m peak positions are shown as orange plus symbols. The SPARKS 1.3-mm continuum emission is traced by the dark grey contours. The dot and dashed arrows show red-shifted and blue-shifted CO outflows reported in Zhang et al. (2019). The lime green arrows represent the ionised jet as identified in Purser et al. (2016).

With a mass of ~9500 $M_\odot$ (Urquhart et al. 2018) this clump is one of the most massive in the ATLASGAL catalogue. The bolometric luminosity of the clump is $1.4 \times 10^5\,L_\odot$, which is sufficient to indicate the presence of a single star of spectral type O6 or a cluster consisting of several high-mass stars where the most massive is of spectral type O7. The presence of multiple high-mass cores and embedded infrared sources including a compact HII region, a high-mass YSO and newly identified HC HII region as well as being associated with numerous EGOs and methanol and OH masers make it clear this clump is hosting a young protocluster.

**G339.681−1.207 (Figure 20):** The radio source G339.681−1.207 is embedded in the ATLASGAL clump AGAL339.681−01.209, which is located at a distance of 2.4 kpc and has a mass and bolometric luminosity of ~600 $M_\odot$ and 1200 $L_\odot$, respectively (Urquhart et al. 2022). The SPARKS data reveals the presence of three cores located towards the centre of the clump, these are labeled as MM1, MM2 and MM3 and have masses of 46, 24 and 21 $M_\odot$, respectively.[3] MM1 and MM2 are located towards the centre of the clump (~2 arcsec) and MM3 is slightly offset from the centre of the clump (~7 arcsec).

The most massive core appears to be relatively quiescent with no known masers or EGO associations. The compact high-frequency radio source is coincident with the lowest mass core (MM3) being located ~2.4 arcsec from the centre of the core, corresponding to a physical offset of 0.03 pc. The Lyman flux for G339.681−1.207 is $9.3 \times 10^{44}$ photons s$^{-1}$ corresponding to a ZAMS star of spectral type B2. The core is associated with a methanol (MMB G339.682−01.207; Caswell 2009) and OH (OH 339.684−01.21) masers, both of which are tightly correlated with the peak of the submillimetre emission (angular offsets are smaller than the positional uncertainties).

The core MM2 is located towards the centre of the ATLASGAL clump and is associated with an methanol maser (MMB G339.681−01.208) and so is likely hosting a young high-mass star. Analysis of the 18 GHz map reveals two radio sources, the bright radio source associated with MM3 discussed in the previous paragraph, and a second weaker radio source (G339.6804−1.2081) with a peak flux of 0.46 mJy and a SNR of 7.1. This source was not detected in the 24 GHz radio map ($f_{24\mathrm{GHz}} < 0.5$ mJy; 3$\sigma$ upper limit) and was considered a low reliability detection and was, therefore, not included in our analysis. However, the tight positional correlation with this radio source with the centre of MM2 and the methanol maser has strengthened it reliability and so we include it in the discussion of this region. The 24 GHz flux upper limit is similar to the 18 GHz flux and so we can assume it is optically thin at both bands. Using this we have determined the Lyman flux to be $3.7 \times 10^{44}$ photons s$^{-1}$ corresponding to a ZAMS star of spectral type B2.

[3] These masses have been re-scaled using a distance of 2.4 kpc and so differ from the values given in Csengeri et al. 2017a, which were determined using a distance of 1.8 kpc.

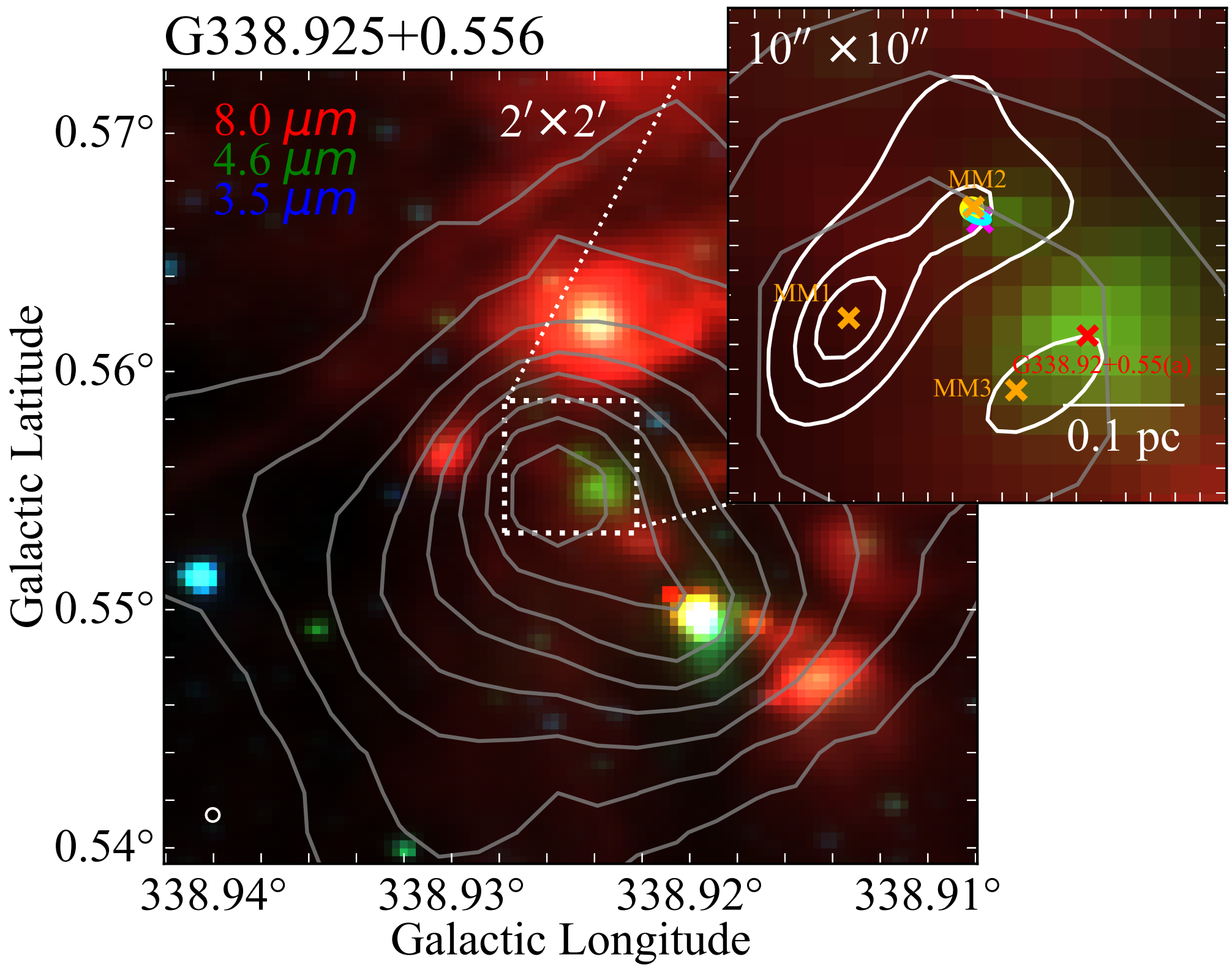

**Figure 19.** Three-colour composite images from GLIMPSE 8-$\mu$m (red), 4.6-$\mu$m (green) and 3.5-$\mu$m (blue) bands for HC H II region G339.884−1.25. The 870-$\mu$m dust emission from ATLASGAL is traced by the grey contours. In the upper right zoomed panel, the cyan ellipse shows the position of the high-frequency (24-GHz) radio source and the magenta cross shows the position of the MMB methanol maser. The position of the OH maser is shown as a yellow circle. The SPARKS 1.3-mm dust continuum emission is traced by white contours and the centres of the three cores identified by Csengeri et al. (2017a) are indicated by orange crosses and labeled MM1, MM2 and MM3 accordingly. The red cross show the peak position of EGOs as identified in Cyganowski et al. (2008).

This high-mass dense clump contains three compact cores. The most massive core is mid-infrared quiet while the two lower mass cores are both associated with embedded H II regions and OH and methanol masers. Similarly to the first clump discussed in this subsection this would appear to be hosting a young protocluster.

The present work allows us to further characterise the nature of the most massive deeply embedded sources of the SPARKS sample. Csengeri et al. 2017b identified these sources as mid-infrared quiet clumps based on mid-infrared emission and luminosity arguments, that could host deeply embedded (proto)stars with masses corresponding to B0.5 or later main sequence stars. Our analysis shows that the most massive deeply embedded objects are B2-B0.5 type main-sequence stars, providing independent confirmation to the initial classification by Csengeri et al. 2017b

## 5 SUMMARY AND CONCLUSIONS

We have carried out high-resolution (~0.5-arcsec) and high-frequency (18−24-GHz) ATCA observations towards a sample of young H II regions that were selected on the basis of being unresolved and optically thick between 5−23.7 GHz with associated methanol masers. The observations were made using the 6A antenna configuration and provide continuum coverage from 17 to 25 GHz. The main aim of this work is to identify and confirm new HC H II regions based on the hypothesis that methanol masers are an excellent tracer of young and embedded high-mass stars. From the 39 observed fields, we successfully identified 27 reliable radio detections above 3 $\sigma$. Utilising various archival multi-wavelength radio-data sources (such as CORNISH-North, CORNISH-South, MAGPIS and THOR), we constructed SEDs between 5 and 24 GHz, where available. These SEDs allowed us to derive physical properties, including size, electron density ($n_e$), emission measure, Lyman continuum photon flux ($N_{\mathrm{Ly}}$), and turnover frequency ($\nu_t$), assuming an ionization-bounded H II region with a uniform electron density model. We have identified 13 HC H II regions, 6 transition objects, 6 UC H II regions and one radio jet candidate based on their physical parameters and mid-infrared environments. The majority of these are newly reported in the work and one source is mid-infrared dark at 24 µm.

Our main findings are as follows:

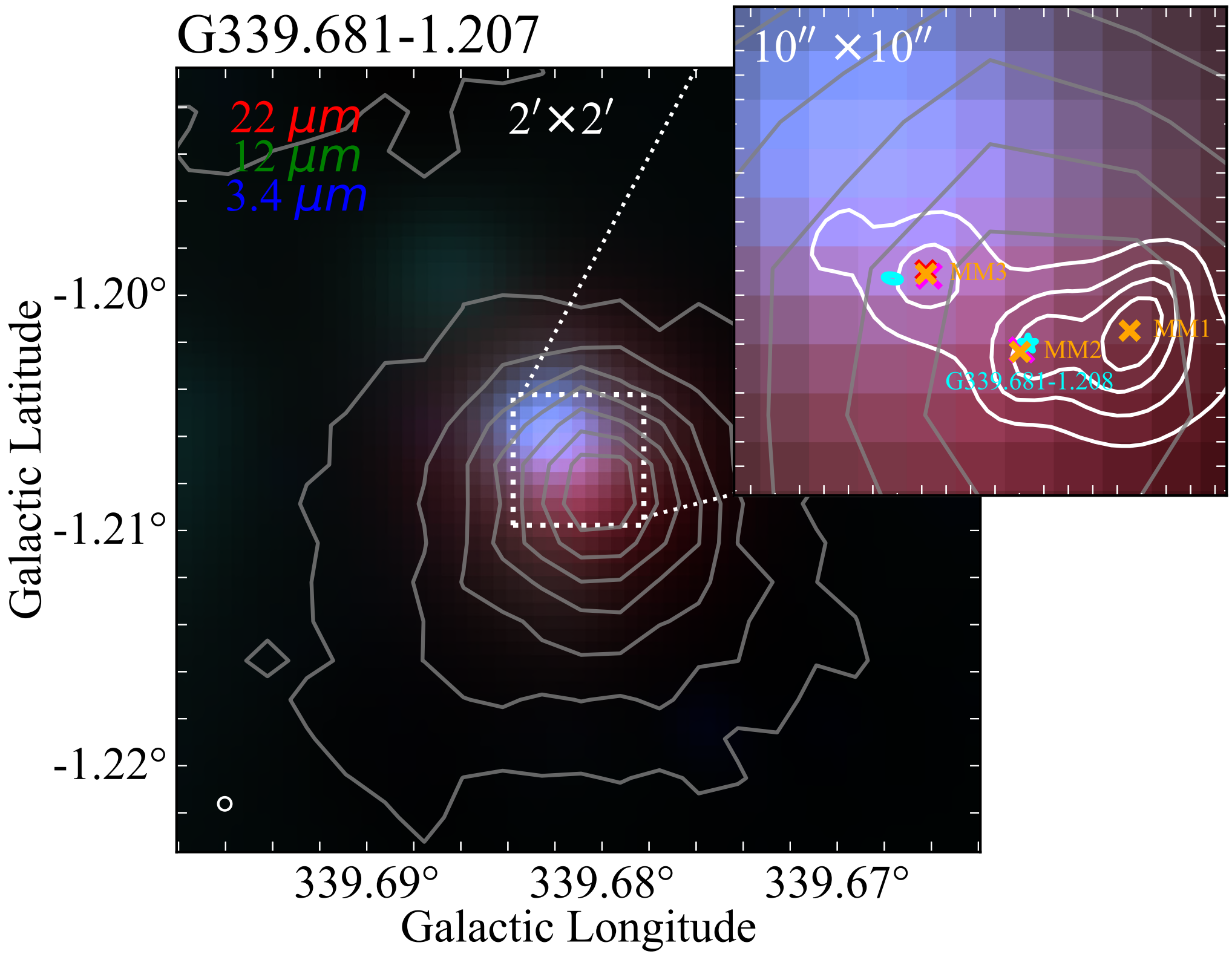

**Figure 20.** Three-colour composite images from WISE 22-$\mu$m (red), 12-$\mu$m (green) and 3.4-$\mu$m (blue) bands for HC Hii region G339.681−1.207. The 870-$\mu$m dust emission from ATLASGAL is traced by the grey contours. In the upper right zoomed panel, the cyan ellipse shows the position of the high-frequency (24-GHz) radio source and the magenta crosses shows the position of the MMB methanol maser. The OH masers is shown as a red cross. The cyan star shows the position of the marginal radio source at 18 GHz. The SPARKS 1.3-mm continuum emission is traced by white contours and the centres of the three cores identified by Csengeri et al. (2017a) are indicated by orange crosses and labeled MM1, MM2 and MM3 accordingly.

- We identify a total of 20 HC Hii regions and intermediate objects located between the class of HC Hii and UC Hii regions, thereby increasing the previously known sample size by ∼80 per cent.

- The youngest Hii regions are deeply embedded with dusty clumps and are found to be associated with a variety of masers, molecular outflows and extended green objects. All of which are excellent probes of the final stages of massive star formation.

- The HC Hii regions identified in this study surround younger and less massive stars. Our observations are specifically targeted towards sites of methanol masers, as they are more sensitive to lower-luminosity sources. Approximately 70 per cent of our sample is associated with a methanol maser within 3 arcsec. 15 per cent of our HC Hii regions do not have a methanol-maser counterpart. Our observational approach targeted towards methanol masers proves to be an effective method for identifying HC Hii regions of massive star formation.

- There are three HC Hii regions and one HC Hii/Jet candidate that remain optically thick at 24 GHz in this study, these sources are considered to be the youngest in our sample. These regions show signs of active accretion within the presence of ionised gas suggesting the protostellar objects exciting the HC Hii regions will ultimately become more massive.

In this study we observed a sample of compact Hii regions selected from 141 methanol masers and successfully identified 14 HCHii regions and 6 transition objects constituting approximately 15 per cent of the total sample. Extrapolating these results to the remaining sample of 831 methanol masers (Caswell 2010), we can anticipate identifying approximately 170 new young and compact Hii regions. While the exact number of detections depend on various factors, this scenario highlights the effectiveness of our observational approach and contributes significantly to increasing the sample of HC Hii regions and to the comprehensive understanding of high-mass star formation.

## ACKNOWLEDGEMENTS

We thank the referee for their comments and suggestions that have improved this work. AYY acknowledges the support from the National Key R&D Program of China No. 2023YFC2206403, and from National Natural Science Foundation of China (NSFC) grants No. 12303031 and No. 11988101. MAT gratefully acknowledges the support of the Science & Technology Facilities Council through grant award ST/W00125X/1.

## DATA AVAILABILITY

The data underlying this article are available in the article and in its online supplementary material. The full version of Table 2 is available on CDS. Copies of the SEDs and 18 and 24 GHz radio maps are available in the Appendix

## APPENDIX A: EXCLUDED RADIO MAPS

In Fig. A1 we show examples of the radio fields that have been excluded from this work. This is a continuation of Fig. 3

## APPENDIX B: RADIO SED

In Fig. B1 we present the SEDs for all the H II regions identified in this work.

## APPENDIX C: MID INFRARED ENVIRONMENTS OF HC H II REGIONS

In Fig. C1 we present three-colour composite images of the mid-infrared environment for the HC H II regions identified in this work.

## APPENDIX D: RADIO MAPS OF ALL DETECTIONS

In Fig. D1 we present the radio maps for all radio detections at 18 and 24 GHz.

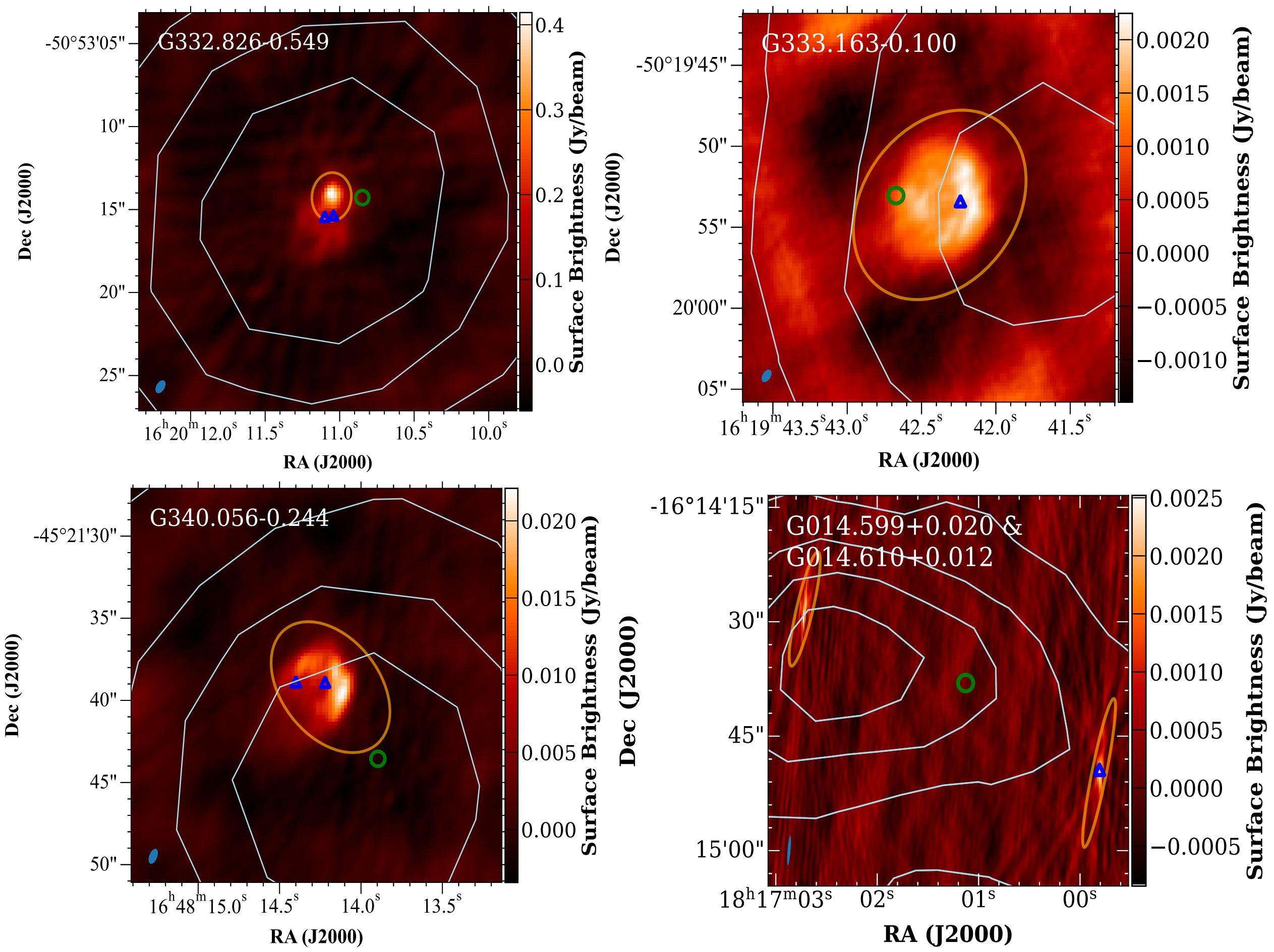

**Figure A1.** Radio maps of the eight excluded fields. The orange ellipse shows the resultant fit to the radio emission while the green circles shows the position of the methanol maser(s) located in the field. The blue triangles show the position of any low-frequency (5 GHz) radio counterparts. The grey contours trace the 870 µm dust emission from ATLASGAL (Schuller et al. 2009). The filled blue ellipse in the bottom left hand corner of each image indicates the size and orientation of the synthesised beam.

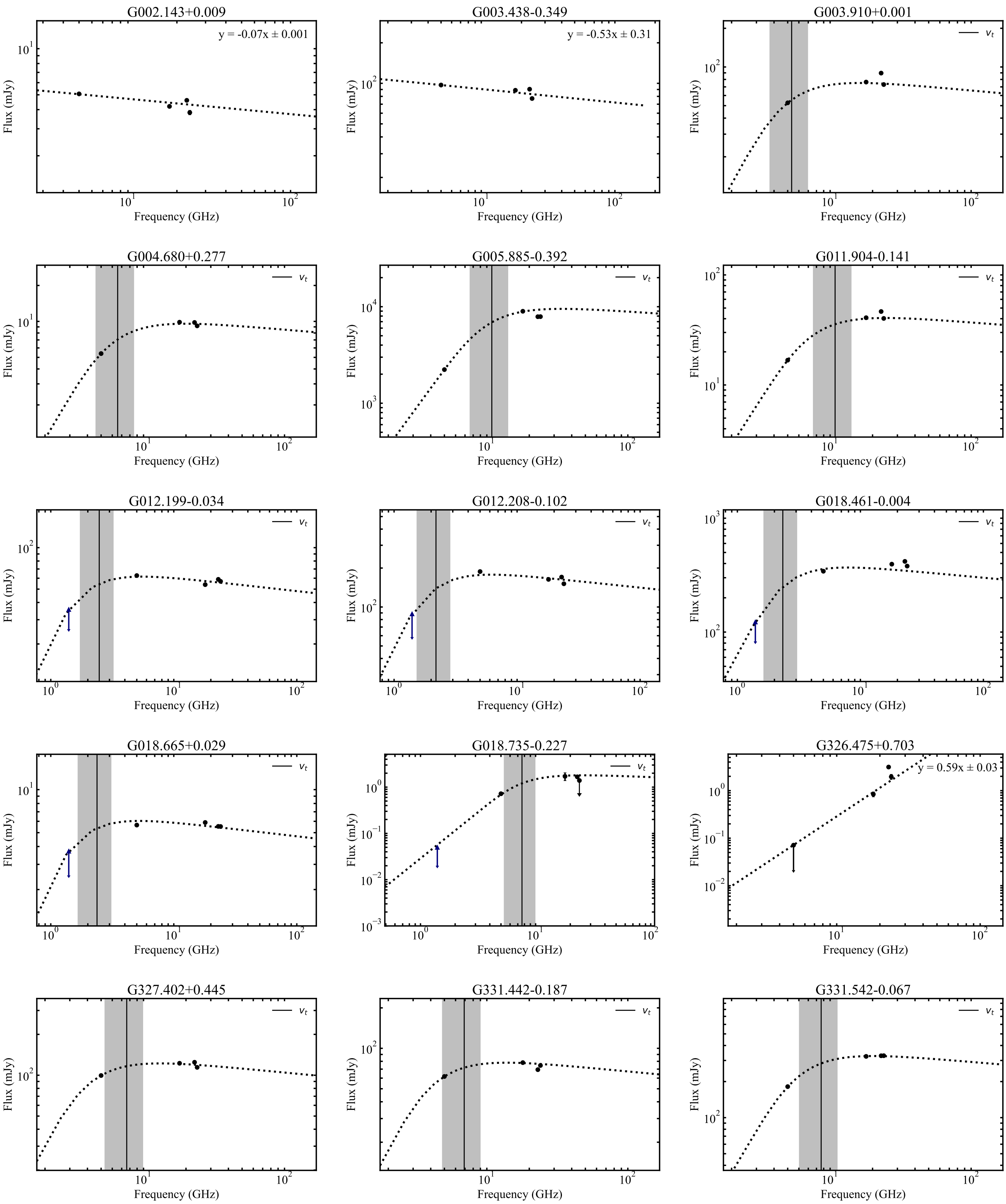

**Figure B1.** Radio SED fittings for the flux densities between 5 − 24 GHz. The solid vertical lines shows where the expected turnover frequency of the object lies. The grey shaded area represents the error associated with the turnover frequency

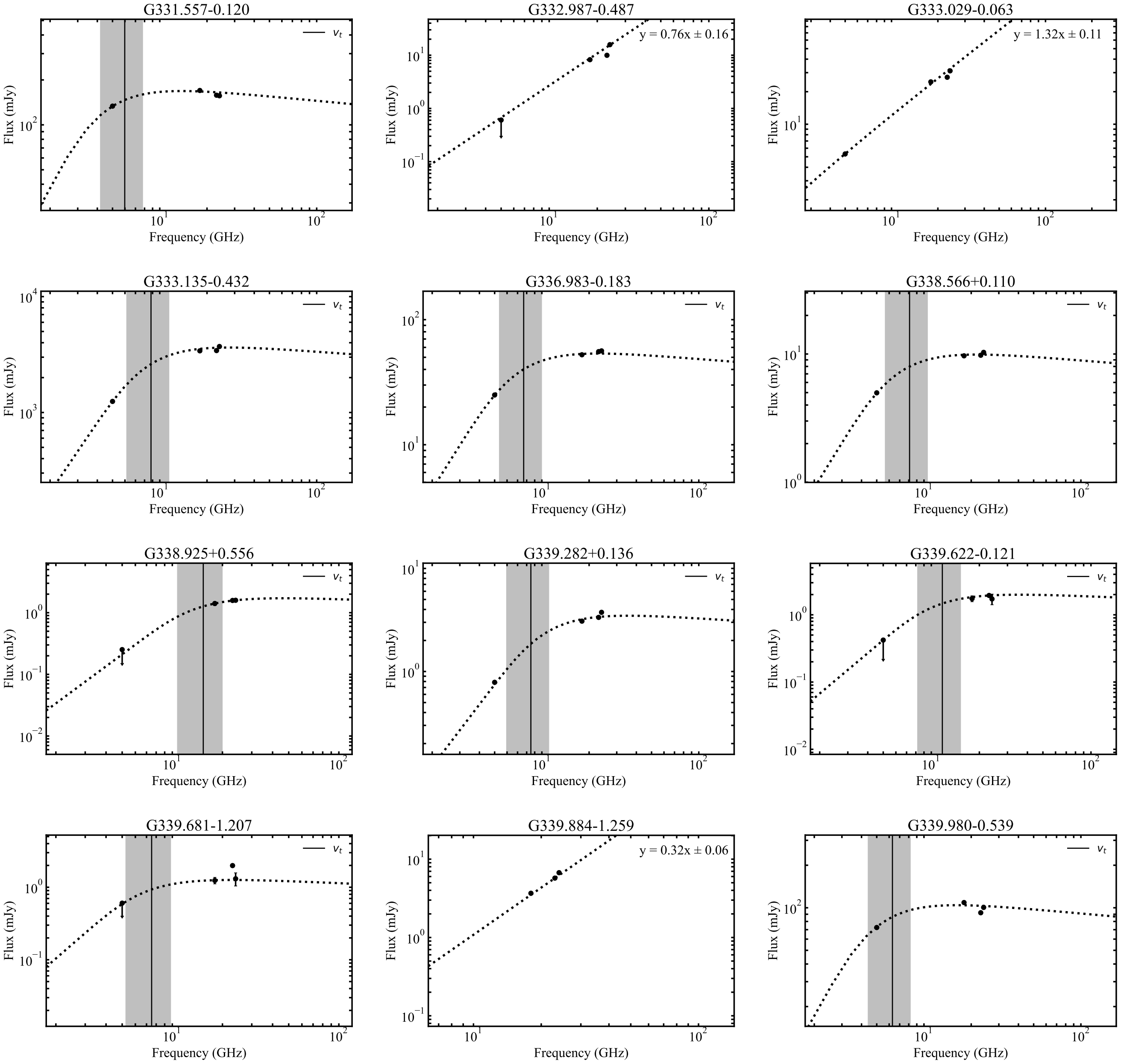

**Figure B1.** Cont.

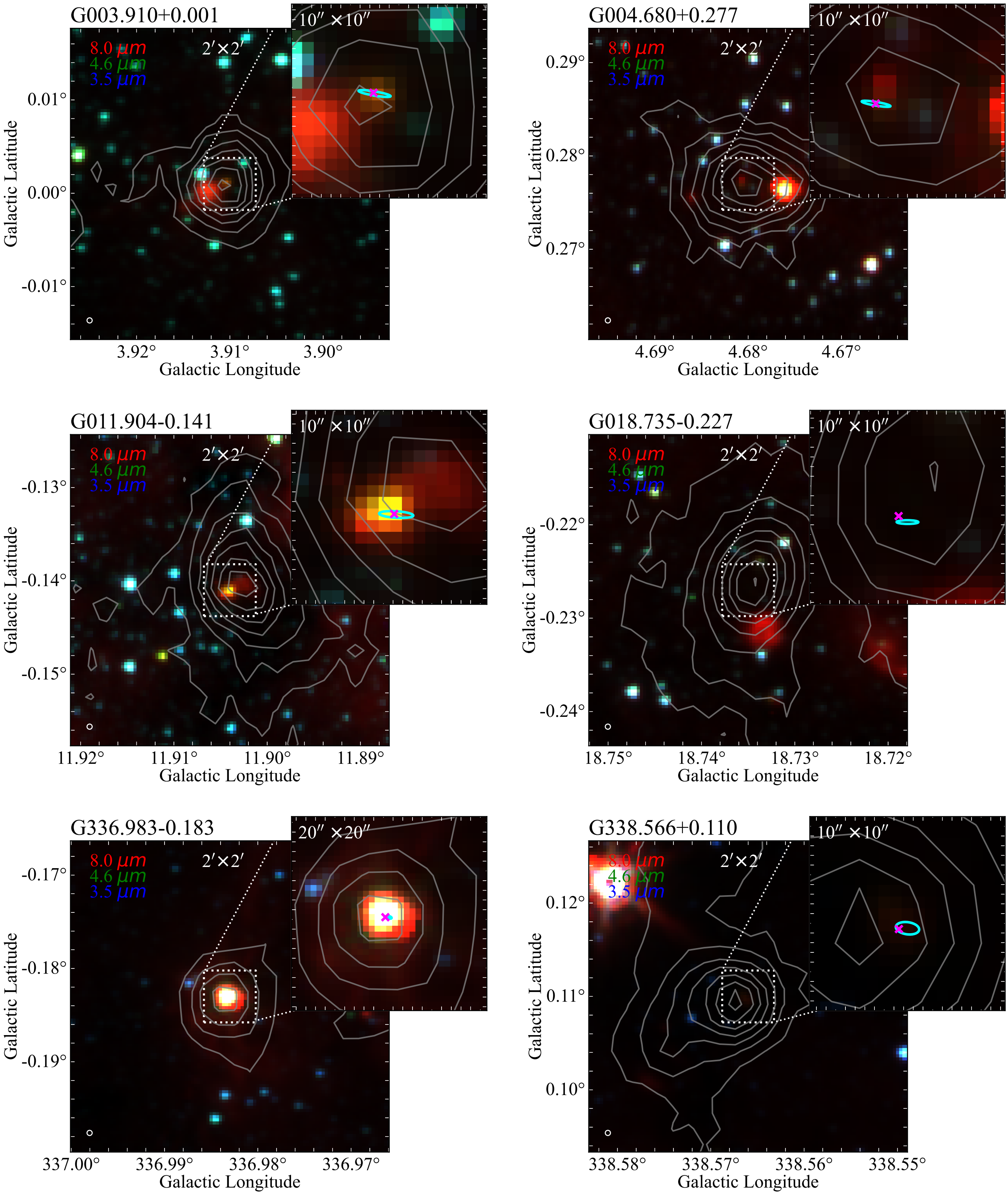

**Figure C1.** Three-colour composite images from Spitzer GLIMPSE 8-$\mu$m (red), 4.5-$\mu$m (green) and 3.6-$\mu$m (blue) bands. The 870-$\mu$m dust emission from ATLASGAL is traced by the grey contours (Schuller et al. 2009). In the upper-right zoomed panel, the cyan ellipse shows the high-frequency (24-GHz) radio source and the magenta cross shows the position of the MMB methanol maser.

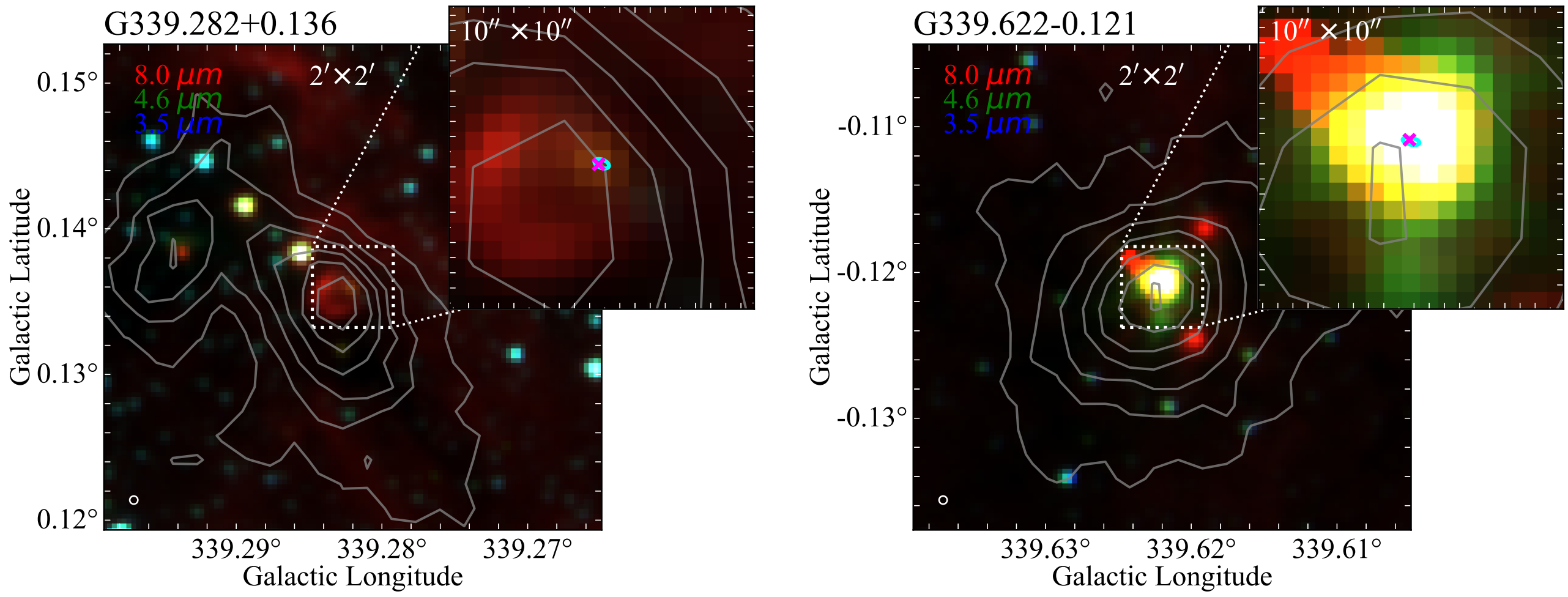

**Figure C1.** Cont.

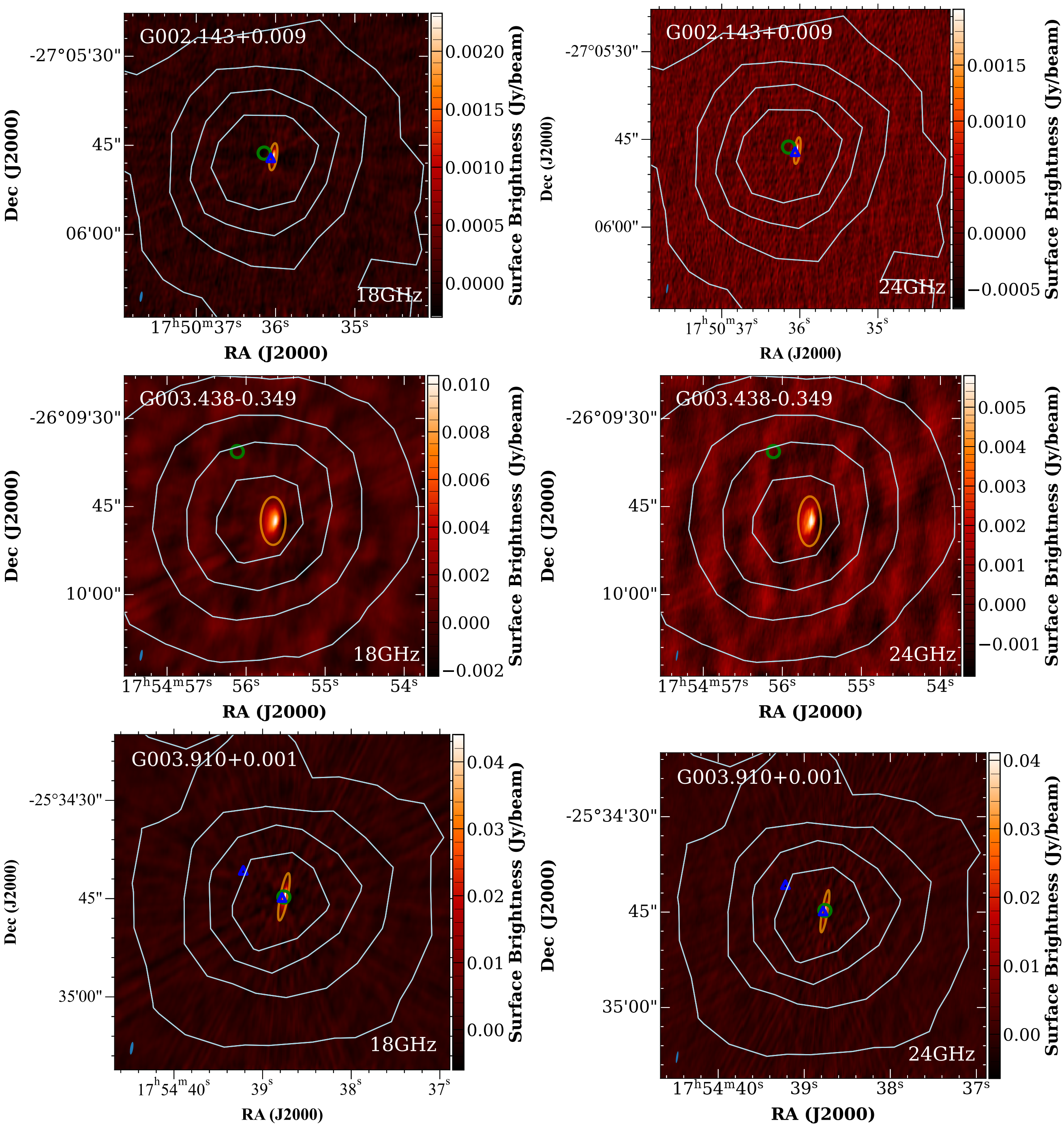

**Figure D1.** Radio maps of all the radio detections at 18 and 24 GHz. The orange ellipse shows the resultant fit to the radio emission while the green circles shows the position of the methanol masers located in the field. The blue triangles show the position of any low-frequency (5 GHz) radio counterparts. The grey contours trace the 870 µm dust emission from ATLASGAL (Schuller et al. 2009). The filled blue ellipse in the bottom left hand corner of each image indicates the size and orientation of the synthesised beam. The radio name is given in the top left hand corner.

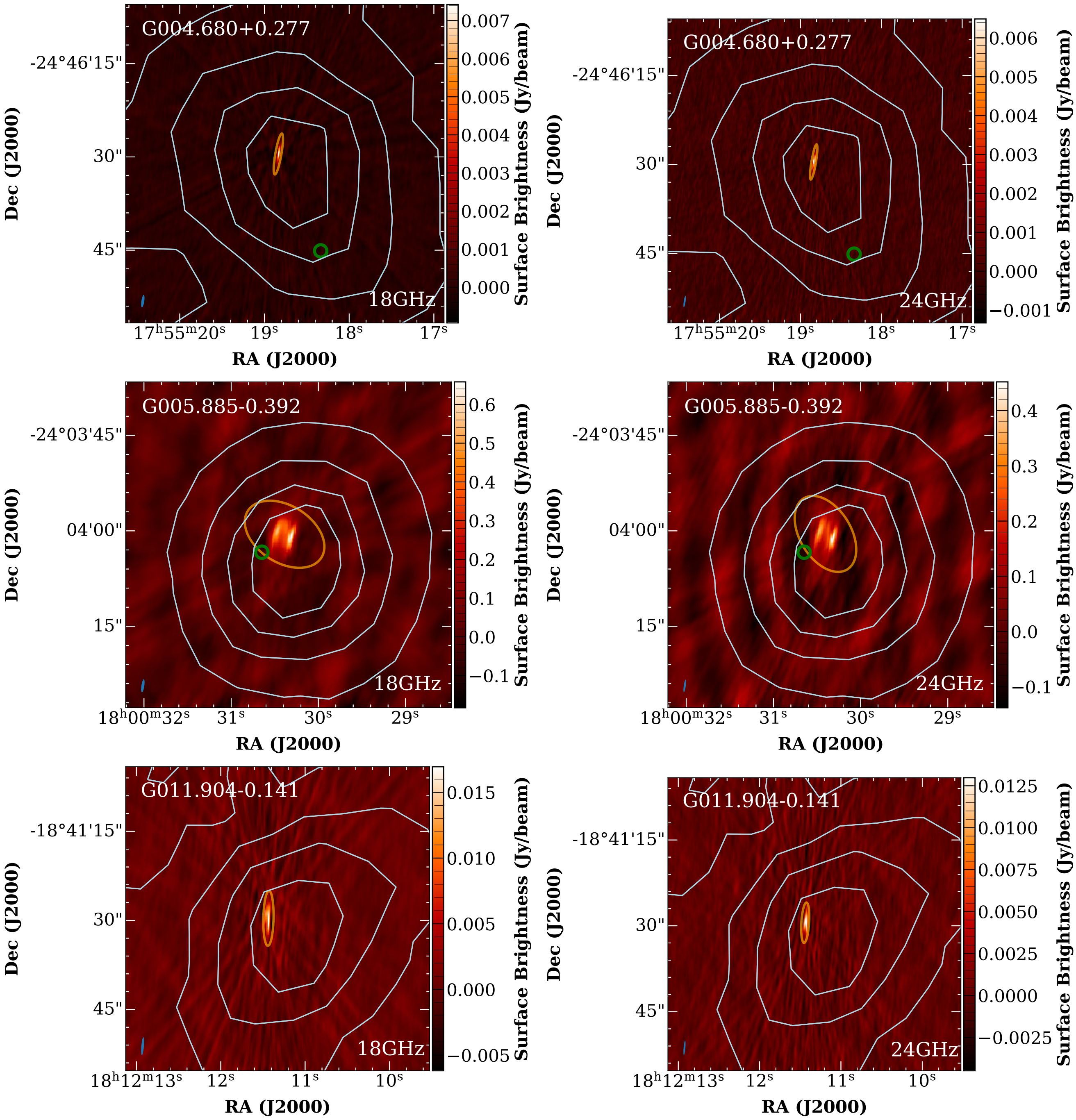

**Figure D1.** Cont.

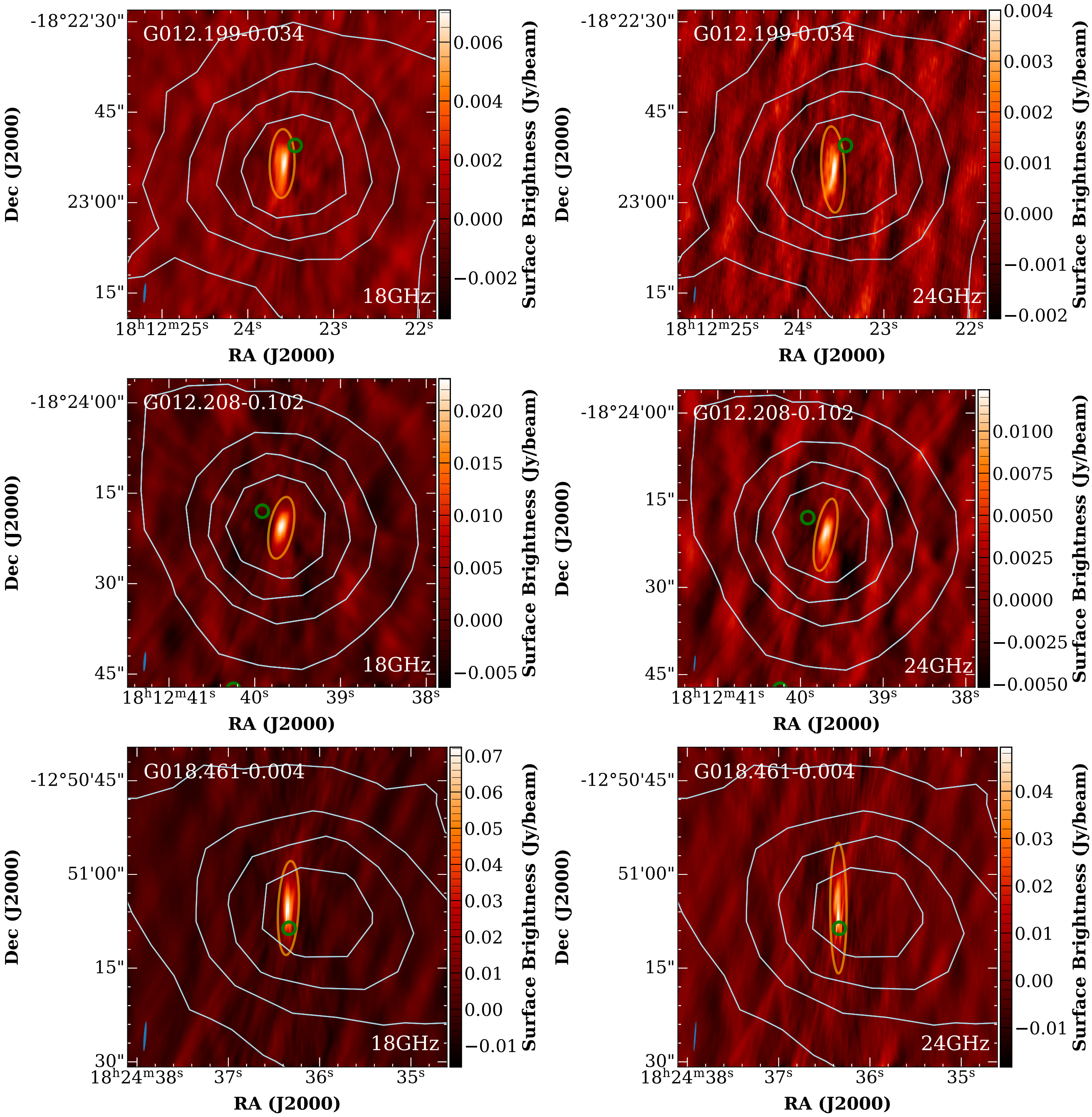

**Figure D1.** Cont.

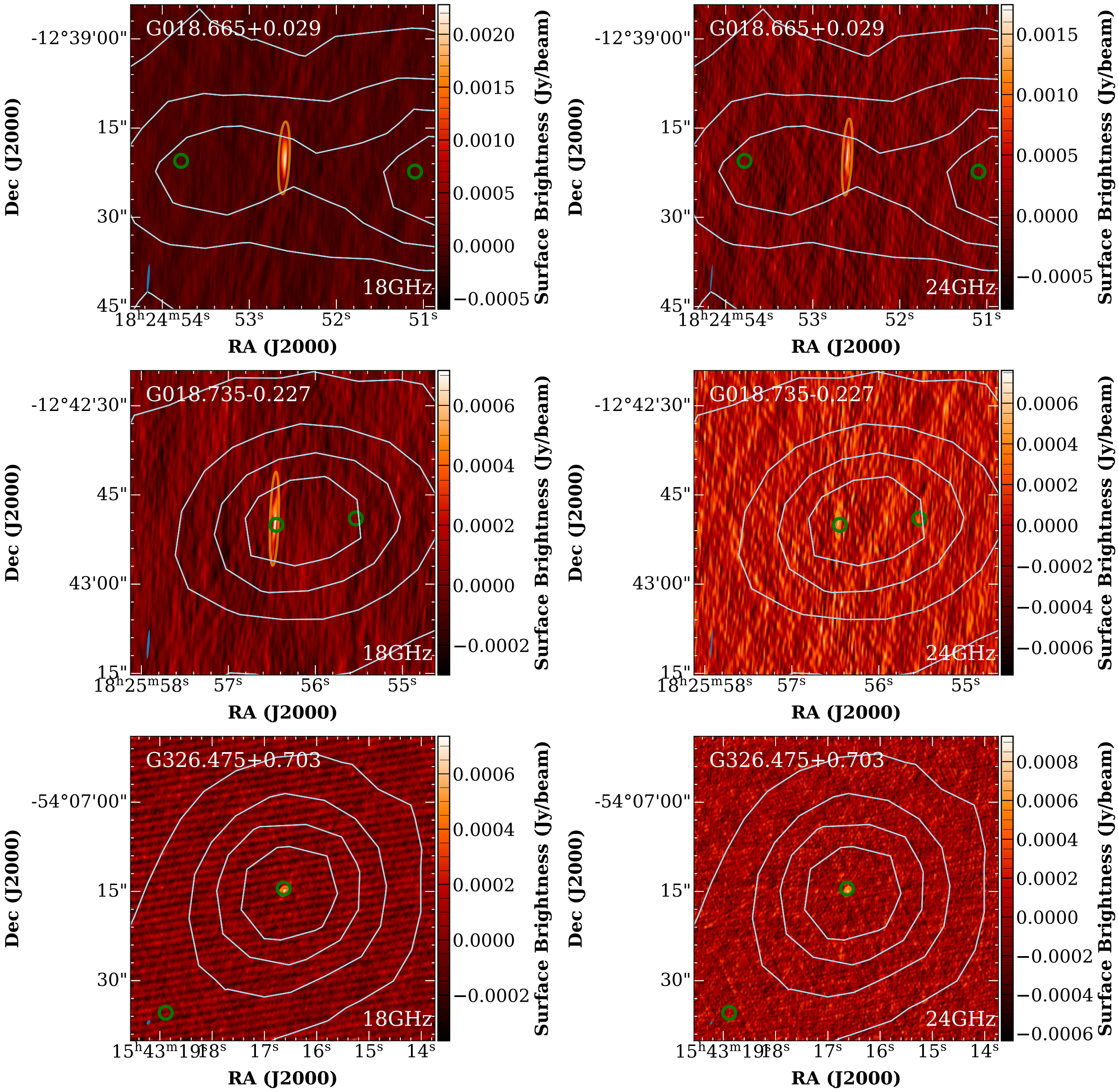

**Figure D1.** Cont.

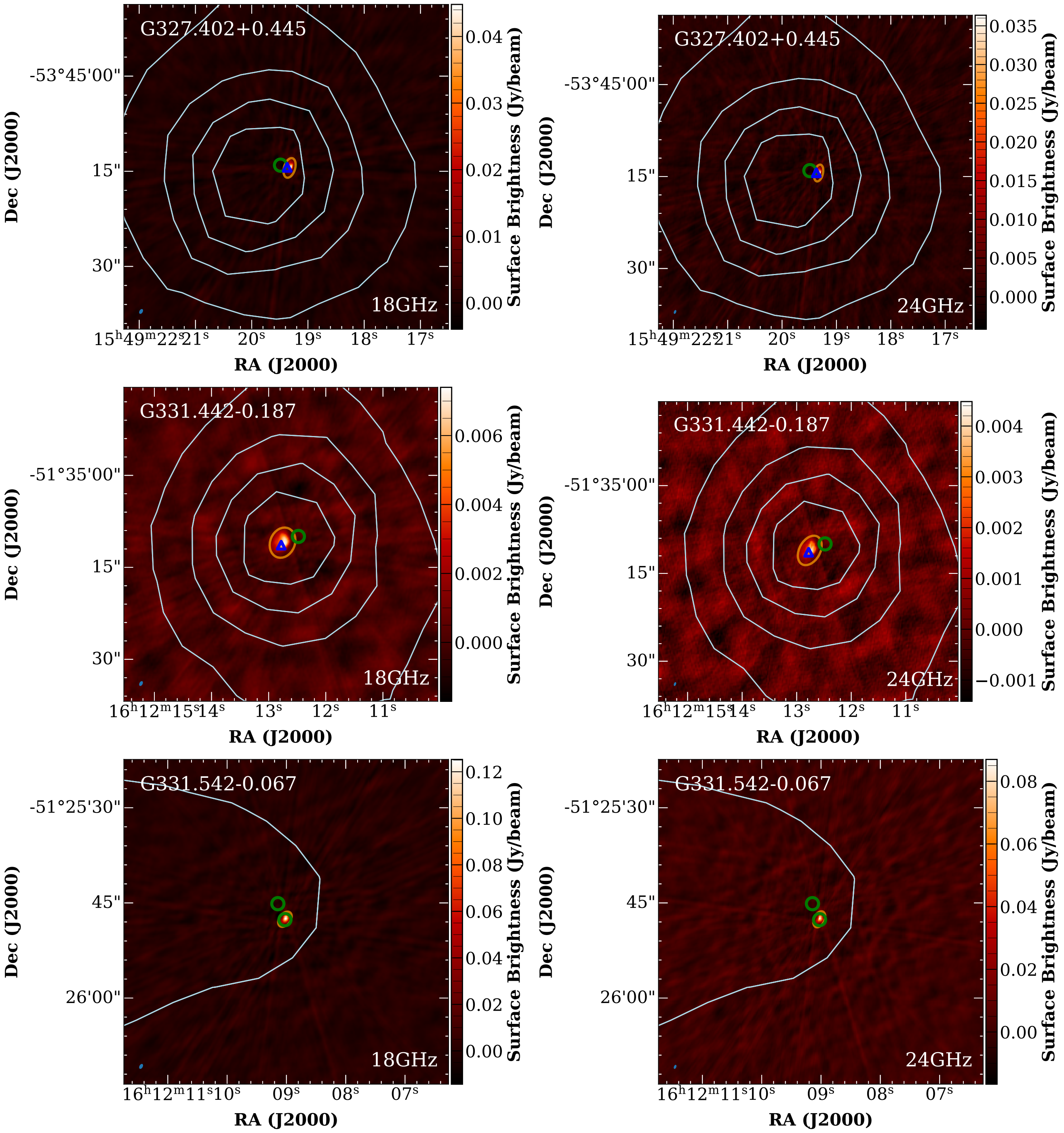

**Figure D1.** Cont.

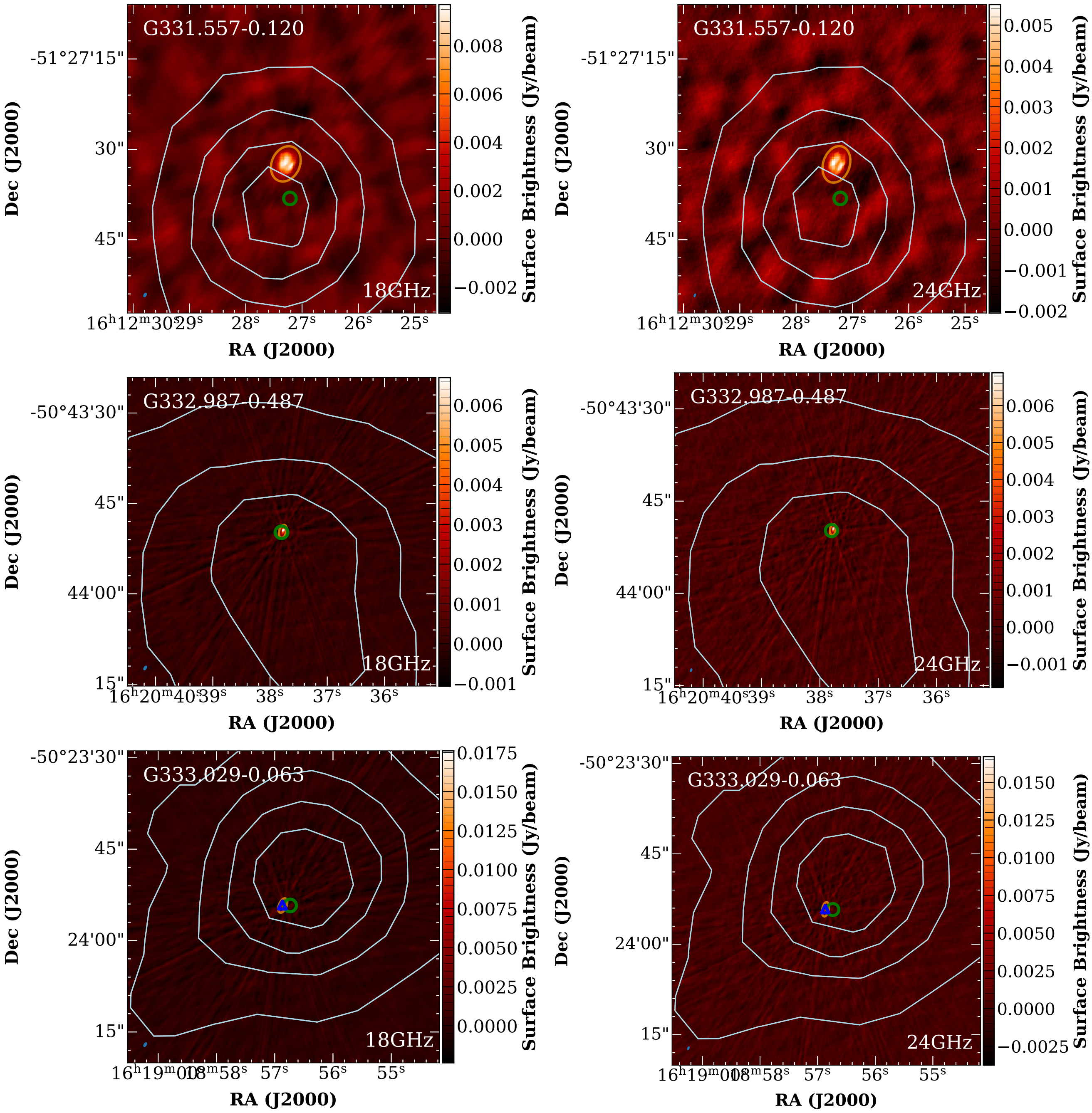

**Figure D1.** Cont.

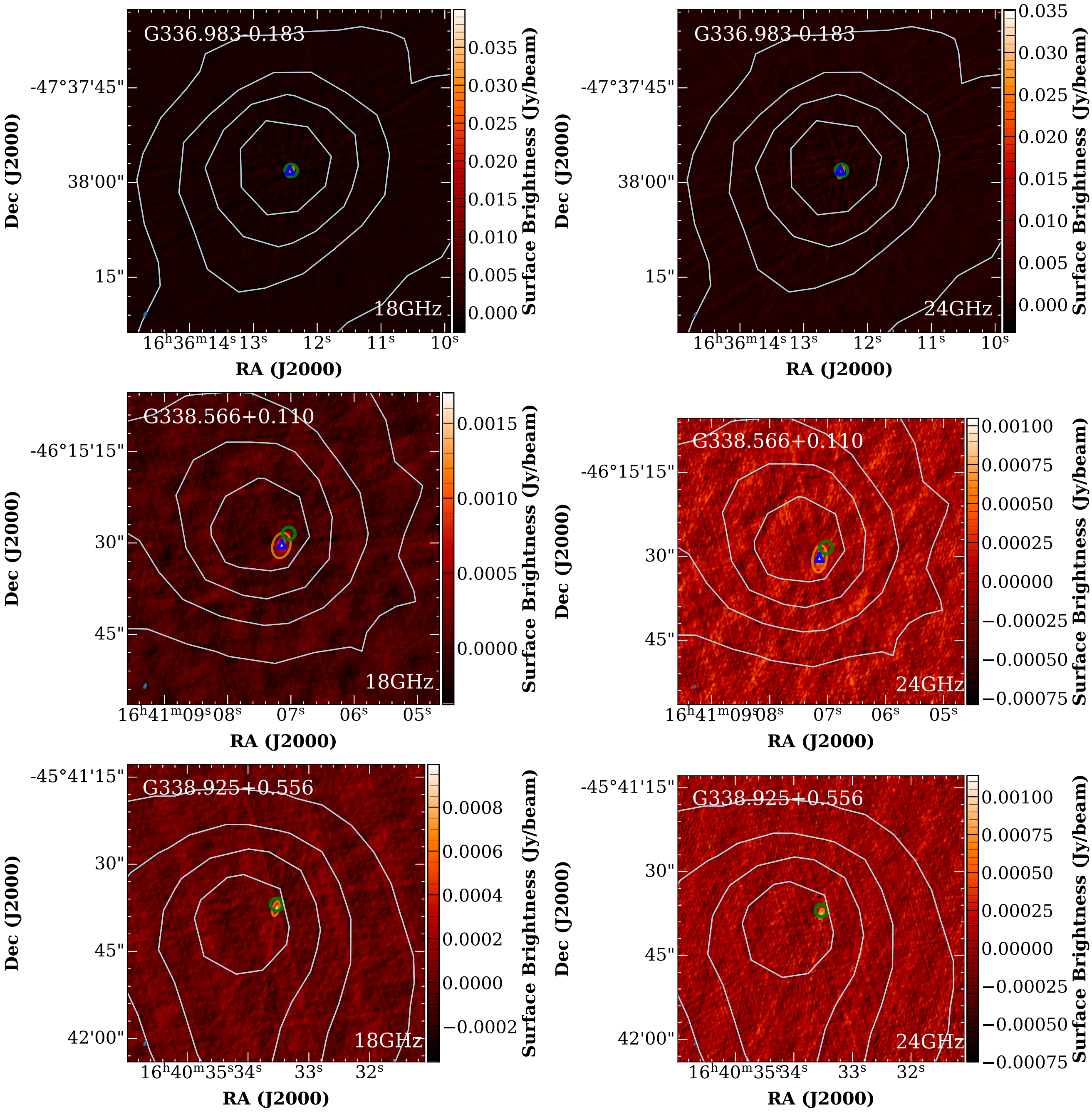

**Figure D1.** Cont.

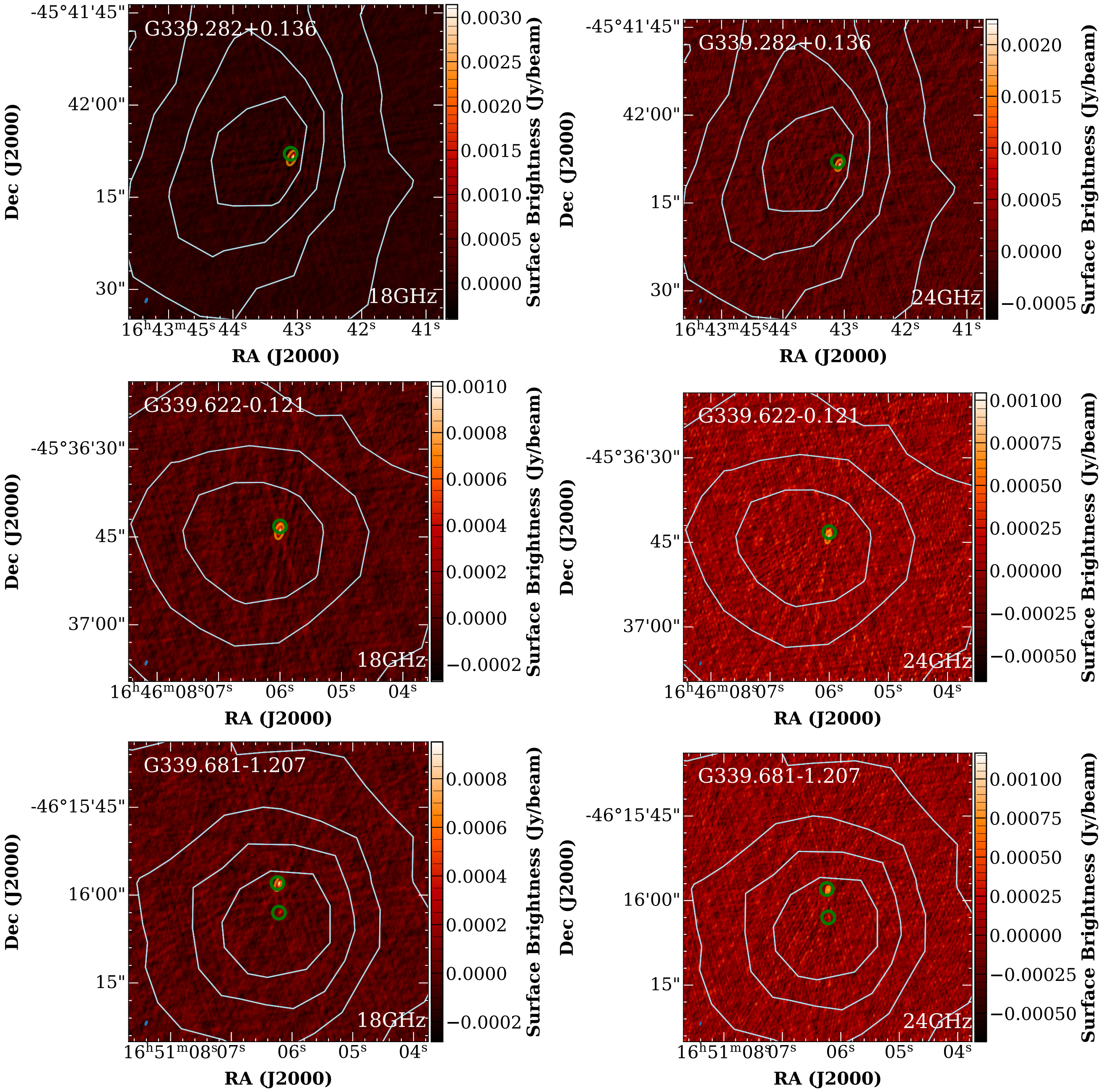

**Figure D1.** Cont.

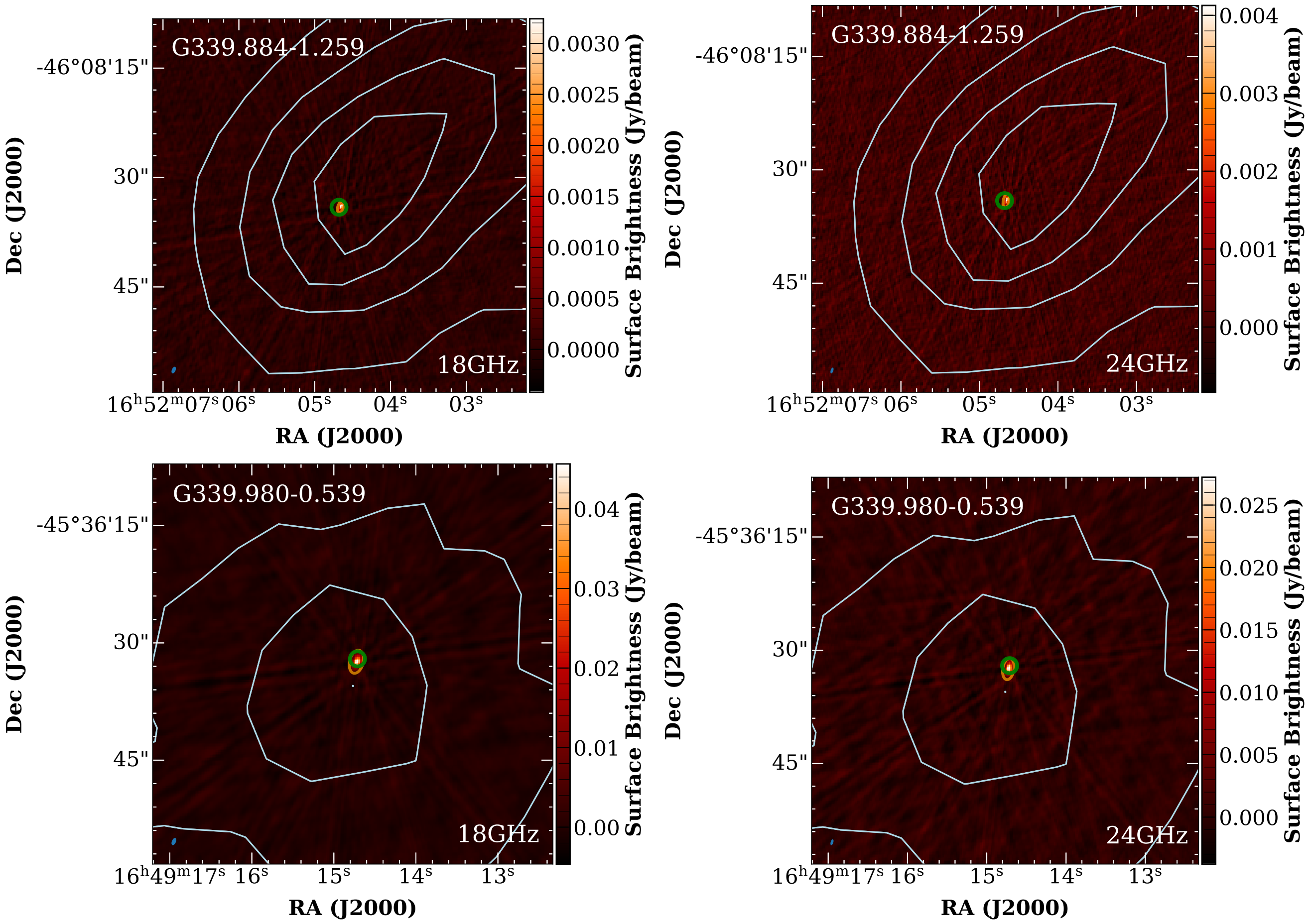

**Figure D1.** Cont.